\documentclass[a4paper,10.5pt]{article}
\usepackage{amsfonts,amsmath,amsbsy,amssymb,amsthm,bm,graphicx,latexsym,graphics,epsfig,subfigure,color,booktabs,caption}
\usepackage{mathrsfs}
\usepackage{longtable,color}
\usepackage{multirow,epsf}
\usepackage{setspace}
\usepackage{float,wasysym}
\usepackage{color}
\usepackage{enumerate} 
\usepackage[dvipsnames]{xcolor}
\usepackage[round, sort]{natbib}
\usepackage[normalem]{ulem}
\usepackage[colorlinks=true,urlcolor=blue,citecolor=blue,linkcolor=blue,bookmarks=true]{hyperref}
\usepackage{tikz}
\usepackage{multicol}
\usepackage[T1]{fontenc}
\usepackage[utf8]{inputenc}

\makeatletter
\newcommand*\rel@kern[1]{\kern#1\dimexpr\macc@kerna}
\newcommand*\widebar[1]{%
  \begingroup
  \def\mathaccent##1##2{%
    \rel@kern{0.8}%
    \overline{\rel@kern{-0.8}\macc@nucleus\rel@kern{0.2}}%
    \rel@kern{-0.2}%
  }%
  \macc@depth\@ne
  \let\math@bgroup\@empty \let\math@egroup\macc@set@skewchar
  \mathsurround\z@ \frozen@everymath{\mathgroup\macc@group\relax}%
  \macc@set@skewchar\relax
  \let\mathaccentV\macc@nested@a
  \macc@nested@a\relax111{#1}%
  \endgroup
}
\makeatother

\allowdisplaybreaks

\newtheorem{The}{Theorem}

\newtheorem{Lem}[The]{Lemma}

\newtheorem{Def}[The]{Definition}
\newtheorem{Exa}[The]{Example}

\numberwithin{The}{section}

\begin{document}
\title{\vskip -1cm \textbf{
On multivariate contribution measures of systemic risk with applications in cryptocurrency market}
}
\author{Limin Wen$^{a}$, Junxue Li$^{a}$, Tong Pu$^{b}$, Yiying Zhang$^{b}$\thanks{Corresponding author. E-mail: zhangyy3@sustech.edu.cn.}
\\
{\footnotesize a. Research Center of Management Science and Engineering, Jiangxi Normal University, Nanchang, 330022, China.}\\
{\footnotesize b. Department of Mathematics,  Southern University of Science and Technology, Shenzhen, 518055, China.}\\}
\date{\today}
\maketitle

\begin{abstract}

Conditional risk measures and their associated risk contribution measures are commonly employed in finance and actuarial science for evaluating systemic risk and quantifying the effects of risk interactions. This paper introduces various types of contribution ratio measures based on the MCoVaR, MCoES, and MMME studied in \cite{ortega2021} and \citet{das2018} to assess the relative effects of a single risk when other risks in a group are in distress. The properties of these contribution risk measures are examined, and sufficient conditions for comparing these measures between two sets of random vectors are established using univariate and multivariate stochastic orders and statistically  dependent notions. Numerical examples are presented to validate these conditions. Finally, a real dataset from the cryptocurrency market is used to analyze the spillover effects through our proposed contribution measures.

\noindent
\\[2mm]
\end{abstract}

\baselineskip17.5pt

\thispagestyle{empty}

\section{Introduction}
In actuarial science and finance, risks are often not isolated events but are highly correlated and capable of spreading. When financial institutions face adverse conditions, such risks can quickly propagate through complex market interconnections, creating a domino effect that escalates losses from a single institution to the entire market, potentially triggering widespread systemic risk. Systemic risk is a core concept in these fields, as it leads to market failures, large-scale financial crises, and profound, long-lasting impacts on the real economy. Typical examples of systemic risk include financial crises, market crashes, bank runs, and contagion effects spreading across multiple industries \citep{divcpinigaitiene2018,zhang2023systemic}.

In this paper, we focus on the ``capital'' type of systemic risk model, which is a method to assess systemic risk losses and their probability of occurrence, and can be characterized by a risk measure\footnote{Systemic risk models can be categorized into the following five types: (i) early warning systems (EWS); (ii) capital; (iii) liquidity; (iv) contagion; and (v) network. For further details, the reader can refer to \citet{silva2017analysis} and \citet{ellis2022systemic}.}. Among widely recognized risk measures, Value-at-Risk (VaR) and Expected Shortfall (ES) have been extensively discussed in the literature \citep{duffie1997,delbaen2000coherent,acerbi2002}.
 However, they fail to effectively quantify systemic risk as they only consider isolated individual economic entities. 
 Hence, there is a need for novel conditional risk (co-risk) measures to quantify and evaluate systemic risk in financial systems. The Conditional Value-at-Risk (CoVaR), introduced by \citet{tobias2016}, measures the VaR of a particular asset under a certain level of systemic stress. \citet{mainik2014} proposed Conditional Expected Shortfall (CoES), which measures the ES of a specific asset or portfolio under a certain level of systemic stress and is defined by the average tail integral of CoVaR. \citet{acharya2017} introduced the concept of Marginal Expected Shortfall (MES) and empirically validated its effectiveness in predicting emerging risks during the 2007-2009 financial crisis. The theoretical properties of these co-risk measures and their applications in finance and insurance can be found in \cite{kritzman2010,asimit2016,feinstein2017,asimit2018systemic,lin2018,duarte2021,liu2021asymptotics,waltz2022vulnerability,yang2024}. However, these co-risk measures can only assess the interaction effect from one entity to another entity, but cannot characterize the absolute or relative spillover effects, which lays the foundation for relevant research on proposing various risk contribution measures. 

 The class of risk contribution measures can be divided into two types: difference-based and ratio-based contribution measures. The former is usually defined as the difference between conditional and unconditional risk measures, while the latter is defined as the ratio between the difference-based contribution measure and the benchmark unconditional risk measure. For example, \citet{tobias2016} introduced the well-known difference-based and ratio-based  contribution measures in terms of CoVaR when the conditional systemic event is taken as VaR at some fixed level. \cite{girardi2013systemic} also defined the difference-based contribution measures (denoted as $\Delta$CoVaR and $\Delta$CoES) based on CoVaR and CoES with different types of conditional events. From the perspective of stochastic orders and dependence structures, \citet{sordo2018} provided sufficient conditions to rank CoVaR, CoES, and their risk contribution measures $\Delta {\rm CoVaR}$ and $\Delta {\rm CoES}$.  \citet{dhaene2022} introduced conditional distortion (CoD) risk measures and the difference-based distortion risk contribution measures ($\Delta {\rm CoD}$), discussing sufficient conditions to rank different bivariate vectors with respect to these measures. Recently, \cite{zhang2024stochastic} introduced several types of ratio-based distortion risk contribution measures ($\Delta^{\rm R} {\rm CoD}$) and examined sufficient conditions for comparing these measures in terms of a new characterization of the convex transform order.

Most of the above-mentioned works only consider a single risk as the systemic risk event, which runs counter to the reality in the financial market that there might be multiple risks collapsing simultaneously. This scenario hinders the usage of the aforementioned systemic risk measures, which calls for the definitions of \textit{multivariate systemic risk measures}. In fact, multivariate risks are attracting increasing attention from many researchers such as \cite{lee2013}, \cite{sun2018}, and \cite{ling2019}. Let ${\bm X} = (X_1,\cdots,X_n)$ represent a portfolio of risks. Here, we assume $X_2,\cdots,X_n$ represent systemic risk, capturing the volatility and uncertainty of the entire system or market, rather than just the impact of any individual asset or event. The definition of multivariate marginal mean excess (in short, MMME) is initially introduced by \cite{das2018} and the asymptotic behavior is studied under suitable conditions within the framework of multivariate regular variation, hidden regular variation and asymptotic tail independence. The multivariate CoVaR (in short, MCoVaR) and multivariate CoES (in short, MCoES) are firstly formally defined in \cite{ortega2021} to quantify the risk of $X_2,\cdots,X_n$ spilled over to $X_1$. Besides, the difference-based contribution risk measures are also introduced. Utilizing MCoVaR, MCoES and MMME,  and their associated difference-based contribution measures (with unconditional risk measure as the benchmark), \cite{ortega2021} investigated sufficient conditions for implementing comparisons between two different sets of multivariate risk vectors.


In this paper, we introduce two types of multivariate risk contribution ratios to reassess the concepts of MCoVaR, MCoES, and MMME. Our research motivation stems from the fact that when we are more concerned with relative contributions rather than absolute contributions of systemic risks, the effectiveness of those contribution measures studied in \cite{ortega2021} is limited. Therefore, this paper introduces the multivariate risk contribution ratio measures to address this issue, where the first type of benchmark measure uses unconditional risk values such as VaR or ES, which are univariate risk measures, and the second type of benchmark measure uses a multivariate joint risk measure based on the median of systemic events, a method commonly used in financial markets. Based on the newly proposed contribution measures, we theoretically analyze sufficient conditions for comparing two distinct multivariate risk portfolios. In particular, for two risk vectors $(X_1, \dots, X_n)$ and $(Y_1, \dots, Y_n)$, the consistency of co-risk measures is examined under different stochastic orders and dependence assumptions. Further, we compute the values of these new risk measures using real-world dataset in the cryptocurrency market, comparing with existing related risk measures, and analyzing the market interactions.  


The remaining sections of this paper are structured as follows. Section \ref{sec:pre} reviews some basic concepts, including univariate and multivariate stochastic orders, copula functions, and some well-known (conditional) risk measures. Section \ref{sec:mulcontri} introduces several new definitions of multivariate systemic risk contribution ratio measures based on MCoVaR, MCoES and MMME, and establishes sufficient conditions to compare two different risk portfolios under these new measures. Section \ref{sec:Empirical} analyzes the risk co-movement effect in the cryptocurrency market by computing and comparing these proposed contribution risk measures. Section \ref{sec:con} concludes the paper. \textcolor{blue}{All proofs and supplementary definitions are provided in the appendix.}

\section{Preliminaries} \label{sec:pre}
Throughout this paper, let $\bm X = (X_1,\dots,X_n)$ and $\bm Y = (Y_1,\dots,Y_n)$ represent two $n$-dimensional random vectors with joint distribution functions (d.f.) denoted by $F$ and $G$, and joint density functions $f({\bm x})$ and $g({\bm x})$, respectively, which will be abbreviated to ``$\bm X\sim F$'' and ``$\bm Y \sim G$''.  Their marginal distributions are denoted by $F_1,\dots,F_n$ and $G_1,\dots,G_n$, which are continuous and have finite expectations. Additionally, their joint survival functions are denoted by $\overline F$ and $\overline G$, that is, $\overline  F ({\bm x})$$ = \mathbb P ({\bm X} > {\bm x})$ and $\overline G({\bm x})= \mathbb P({\bm Y} >{\bm x})$ for ${\bm x} \in \mathbb R^n$. Let ${\bm x}=(x_1,\dots,x_n)$ and ${\bm y}=(y_1,\dots,y_n)$ be two real-valued vectors in $\mathbb R^n$, we denote ${\bm x} \vee {\bm y} = \left({\rm min}\{x_1,y_1\},\dots,{\rm min}\{x_n,y_n\} \right)$ and ${\bm x} \wedge {\bm y} = \left({\rm max}\{x_1,y_1\},\dots,{\rm max}\{x_n,y_n\} \right)$. 

\subsection{Stochastic orders}
The quantile function for a random variable \( X \) with distribution function \( F_X \) is defined as:
\begin{equation*}
\mathrm{VaR}_{p}(X):={F_X^{ - 1}}(p) = \inf \{ x \in {\mathbb R}|F_X(x) \ge p\},~~ p \in (0,1).
\end{equation*}
Now, we present several pertinent definitions of univariate stochastic orders that will be utilized in subsequent discussions. 

\begin{Def}\label{def:univariate}
Let $X$ and $Y$ be two random variables with distribution functions $F_X$ and $F_Y$, density functions $f_X$ and $f_Y$, and survival functions $\overline{F}_X$ and $\overline{F}_Y$, respectively. Then $X$ is said to be smaller than $Y$ in the: 
\begin{enumerate}[(i)]
    \item usual stochastic order (denoted by $ X \leq_{\rm st} Y$) if  ${\overline F}_X(t) \leq {\overline F}_Y(t)$ for all $t \in {\mathbb R}$;
    \item excess wealth order (denoted by $X\leq_{\rm ew} Y$) if $\mathbb{E}\left[\left( X - F_X^{-1}(p) \right)_+ \right]\le \mathbb{E}\left[\left( Y - F_Y^{-1}(p) \right)_+\right]$ for all $0 < p < 1$, where $x_+={\rm max}(0,x)$;
    \item star order (denoted by $ X\leq_{\star} Y$) if  $F_Y^{-1}(p)/F_X^{-1}(p)$ is increasing in $p \in (0,1)$, \textcolor{blue}{when the two random variables are non-negative};
    \item  expected proportional shortfall order (denoted by $X\leq_{\rm ps} Y$) if
${\rm EPS}_p(X) \leq {\rm EPS}_p(Y)$ for all $p \in D_X\cap D_Y$, \textcolor{blue}{when the two random variables are non-negative}, where $D_X = \{p\in (0,1): F_X^{-1}(p)>0\}$, $D_Y = \{p\in (0,1): F_Y^{-1}(p)>0\}$. Here,
\begin{equation*}
\mathrm{EPS}_{p}(X)=\mathbb{E}\left[ \left( \frac{X-\mathrm{VaR}_{p}(X)}{%
\mathrm{VaR}_{p}(X)}\right) _{+}\right] ~\text{ and }~\mathrm{EPS}_{p}(Y)=%
\mathbb{E}\left[ \left( \frac{Y-\mathrm{VaR}_{p}(Y)}{\mathrm{VaR}_{p}(Y)}%
\right) _{+}\right] .
\end{equation*}
\end{enumerate}
\end{Def}

\textcolor{blue}{In Definition \ref{def:univariate}, (i)–(iii) are referenced in \citet{shaked2007}, and (iv) can be found in \citet{belzunce2012}. It is also known that both of the star order and the expected proportional shortfall order are scaled invariant and the former implies the latter. Interested readers can refer to the monographs \citet{shaked2007} and \citet{belzunce2015introduction} for more detailed discussions.}

Next, for the random vector  $\bm X = (X_1,\dots,X_n)$, we introduce some multivariate  stochastic orders and dependence notions, indicating that in some stochastic sense, larger values of one random vector are associated with larger or smaller values of another random vector.

\begin{Def}  \citep{shaked2007}
Let $\bm X = (X_1,\dots,X_n)$ and $\bm Y = (Y_1,\dots,Y_n)$ be two random vectors with  joint distribution functions $F$ and $G$, respectively. Then
\begin{enumerate}[(i)]
\item the random variable $\{X_i, i\in A^c\}$, is said to be right-tail increasing (decreasing)  in $\{X_j,j \in A\}$, (denoted by $\{X_i, i \in A^c\}\uparrow_{\rm  RTI[RTD]}  \{X_j,j \in A\}$)  if  $\mathbb P(X_i > x_i, i \in A^c| X_j > x_j, j \in A)$ increases (decreases) in $x_j$, where $A$ is a subset of  $\{1,\dots, n\}$ with at least one element, and $A^c$ denotes the complement of $A$;
\item the random vector $\bm X$ is said to be smaller than $\bm Y$ in the  multivariate hazard rate order (denoted by $\bm X \leq_{\rm hr} \bm Y$)  if $\overline F({\bm x}) \overline G({\bm y}) \leq \overline F({\bm x} \wedge {\bm y}) \overline G({\bm x} \vee {\bm y})$ for all ${\bm x}, {\bm y} \in {\mathbb R}^n$; %
\item the random vector $\bm X$ is said to be smaller than $\bm Y$ in the weak multivariate hazard rate order (denoted by $\bm X \leq_{\rm whr} \bm Y$) if  $\overline G (\bm x)/ \overline{F}(\bm x)$ is increasing in $\bm x \in \left\{\bm x: \overline {G}({\bm x}) >0 \right\}$; %
\item the random vector $\bm X$ is said to be smaller than $\bm Y$ in the  usual stochastic order (denoted by $\bm X \leq_{\rm st} \bm Y$)  if  $\mathbb E [h(\bm X)] \leq \mathbb E [h(\bm Y)]$ for all bounded increasing $h: \mathbb R^n \rightarrow \mathbb R$; %
\item the random vector $\bm X$ is said to be multivariate totally positive of order 2 (denoted by ${\rm MTP}_2$)  if $f(\bm x) f(\bm y) \leq f(\bm x \wedge \bm y) f(\bm x \vee \bm y)$ holds for all $\bm x,\bm y \in {\mathbb R}^n$; %
\item the  random vector $\bm X$ is said to be smaller than $\bm Y$ in the multivariate likelihood ratio order (denoted by $\bm X \leq_{\rm lr} \bm Y$) if  $f({\bm x}) g({\bm y}) \leq f({\bm x} \wedge {\bm y}) g({\bm x} \vee {\bm y})$ for all ${\bm x}, {\bm y} \in {\mathbb R}^n$; %
\item the random vector $\hat{\bm X}_i = (X_1,\dots,X_{i-1},X_{i+1},\dots,X_n)$ is said to be stochastically increasing in $X_i$ (denoted by $\hat{\bm X}_i \uparrow_{\rm SI} X_i$) if the conditional distribution $\left\{(X_1,\dots,X_{i-1},X_{i+1},\dots,X_n|X_i = x_i) \right\}$ is stochatically increasing as $x_i$ increases;
\item the random vector $\bm X$ is said to be positive dependent through the stochastic order (or PDS) if $\hat{\bm X}_i \uparrow_{\rm SI} X_i$ for $i \in \{1,\dots,n\}$.
\end{enumerate}
\end{Def}
According to \citet{hu2003}, the following relationships hold:
$$\bm X \leq_{\rm lr} \bm Y \Longrightarrow \bm X \leq_{\rm hr} \bm Y \Longrightarrow \bm X \leq_{\rm whr} \bm Y~~\text{and}~~\bm X \leq_{\rm lr} \bm Y \Longrightarrow \bm X \leq_{\rm st} \bm Y.$$

\subsection{Copula}
Let \( F \) be the joint distribution function of the random vector \( \bm X \) with  continuous marginal distribution functions \( F_1, \dots, F_n \). Then, there exists an $n$-dimensional copula function $C(p_1,\dots,p_n)$ defined on $[0,1]^n$ such that
\[
F(x_1, \dots, x_n) = C(F_1(x_1), \dots, F_n(x_n)), \quad \forall~ x_1, \dots, x_n \in \mathbb R.
\]
Here, the copula function \( C \) captures the dependence structure of the random vector \( (X_1, \dots, X_n) \). Let \( U_i = F_i(X_i) \), which follows a uniform distribution \( U[0,1] \). Then the copula function \( C \) can be re-expressed as:
\[
C(p_1, \dots, p_n) = \mathbb{P}(U_1 \leq p_1, \dots, U_n \leq p_n),
\]
where \( p_i = F_i(x_i) \) for \( i = 1, \dots, n \).
Clearly, it is deduced that
\begin{equation*}
C(p_{1},\dots,p_{n})=F\left( F_{1}^{-1}(p_{1}),\dots
,F_{n}^{-1}(p_{n})\right).
\end{equation*}

The joint tail function, denoted as $\overline{C}$, is expressed as
\begin{equation*}
\overline{C}(p_{1},\dots,p_{n})=\mathbb{P}(X_1>F_{1}^{-1}(p_{1}),\dots,X_n>F_{n}^{-1}(p_{n})).
\end{equation*}
For an $n$-dimensional uniform random vector, the \textcolor{blue}{joint tail function} $\overline C$ can be represented in terms of the copula $C$ as follows \textcolor{blue}{(see Theorem 4.7 in \citet{cherubini2004copula}):
\begin{equation*}
    \overline C(p_1,\dots,p_n) = \sum_{i=0}^n \left[ (-1)^i \sum_{{\bm w}(\bm p) \in Z (n-i,n,1)} C({\bm w (\bm p)}) \right],
\end{equation*}
where $Z(n-i,n,i)$ is the set of the $\binom{n}{i}$ possible vectors with $n-i$ components equal to 1, $i$ components  equal to $p_i$.} 

The Archimedean copula, a prevalent category within the family of copulas, is characterized by a generating function known as the Archimedean generator. The expression for an $n$-dimensional Archimedean copula is given by: 
\begin{equation*}
    C_\psi (u_1, u_2, \ldots, u_n) = \psi^{-1} \left( \psi(u_1) + \psi(u_2) + \cdots + \psi(u_n) \right),
\end{equation*}
where $\psi$ is a strictly decreasing function called the generating function, with its inverse denoted as $\psi^{-1}$. Prominent examples of Archimedean copulas include the Clayton, Gumbel, and Frank copulas, each employing distinct generating functions to model the dependencies among random variables. \textcolor{blue}{For these specific forms of Archimedean copulas, please refer to Appendix \ref{app:copula}.}

Next, the definition of concordance order is provided describing one copula is more positively dependent than the other.
\begin{Def} \citep{nelsen2006}
    Given two $n$-dimensional copulas $C$ and $C'$,  $C$ is smaller than $C'$ in the concordance order (denote by $C \leq_{\rm c} C'$) if $C(\bm p) \leq C'(\bm p)$, for all $\bm p \in [0,1]^n$.
\end{Def}

For copula functions $C$ and $C'$, there also exists a stronger ranking relationship in terms of the weak multivariate hazard rate order \citep{hu2003}, which is defined as follows.
\begin{Def}
 Given two $n$-dimensional copulas $C$ and $C'$, $C$ is said to be smaller than $C'$ in the weak multivariate hazard rate order (denote by $C \leq_{\rm whr} C'$) if  $\overline C'(\bm p)/\overline C(\bm p)$ is increasing in $\bm p \in \left\{\bm p \in [0,1]^n:\overline C(\bm p)>0 \right\}$.
\end{Def}

\subsection{Multivariate co-risk measures}
For an individual risk $X$ with distribution function $F_X$, 
the Expected Shortfall (ES) of $X$ at a given probability level $p\in(0,1)$ is defined as
\begin{equation*}
{\rm ES}_{p}[X] = \frac{1}{1-p} \int_{p}^{1} {{\rm VaR}_t}[X] dt.
\end{equation*}
Essentially, VaR represents the one-sided critical value of asset value loss over a certain holding period at a given confidence level, practically manifesting as an amount serving as the threshold. Compared to VaR, ES considers the magnitude of losses beyond the VaR threshold, making it a more comprehensive measure of risk. ES is particularly suitable when tail risk is of concern or when a more comprehensive risk assessment is needed. Besides, according to the Basel IV accords, the internal/advanced model approach is revised by replacing the VaR measure with the ES measure, which further highlights the importance of ES in solvency regulation; see \cite{kou2013external} and \cite{zaevski2023basel}.


In finance, the interconnections among entities' (e.g. banks or financial institutions) risks can lead to varying levels of systemic risk. To investigate the risk spillover of other individuals on one concerned  entity, \citet{ortega2021} introduced a multivariate co-risk measure called  MCoVaR as follows:
\begin{equation*}
\mathrm{MCoVaR}_{\bm p }[X_{1}|X_{2},\ldots,X_{n}]= {\rm VaR}_{p_1} \left[ X_1 \bigg| \bigcap\limits_{j = 2}^n \left\{{X_j} > {\rm VaR}_{p_j}[X_j]\right\}  \right],
\end{equation*}
where $\bm p =(p_1,\dots,p_n)\in (0,1)^n$. Clearly, MCoVaR is the VaR of the conditional distribution of $X_1$ at level $p_1$, given the joint systemic risk event $\{X_2 > {\rm VaR}_{p_2}[X_2], ..., X_n > {\rm VaR}_{p_n} [X_n]\}$. As a direct generalization, the MCoES is further introduced in \citet{ortega2021} as follows:
\begin{equation*}
\mathrm{MCoES}_{\bm p }[X_{1}|X_{2},\dots,X_{n}]= \frac{1}{1-p_1} \int_{p_1}^1 {\rm MCoVaR}_{t,p_2,\dots,p_n}[X_{1}|X_{2},\dots,X_{n}] dt.
\end{equation*}

Following \citet{das2018}, the MMME is delineated as:
\begin{equation*}
\mathrm{MMME}_{\bm p_{[-1]} }[X_{1}|X_{2},\dots,X_{n}]= \mathbb E\left[ \left(X_1-A_{X,p_{[-1]}}\right)_+ \bigg| \bigcap\limits_{j = 2}^n \left\{{X_j} > {\rm VaR}_{p_j}[X_j]\right\} \right],
\end{equation*}
where ${\bm p}_{[-1]} = (p_2,\dots,p_n) \in (0,1)^{n-1}$, $A_{X,p_{[-1]}} = \sum_{i=2}^{n} a_i {\rm VaR}_{p_i}[X_i]$ and $a_i \in [0,1]$ satisfies  $\sum_{i=2}^{n} a_i=1$.  The $\mathrm{MMME}_{\bm p_{[-1]} }[X_{1}|X_{2},\dots,X_{n}]$ represents the expected excess of \(X_1\) over a threshold \(A_{X, p_{[-1]}}\), conditional on the event that each \(X_j\) exceeds its VaR at level \(p_j\), for \(j = 2, \ldots, n\). The threshold \(A_{X, p_{[-1]}}\) is a weighted sum of the VaRs of \(X_2\) to \(X_n\). This measure captures the expected amount by which \(X_1\) exceeds the threshold, reflecting the marginal mean excess risk under the given joint conditions.

\section{Multivariate conditional risk contribution ratio measures and comparison  results}\label{sec:mulcontri}
\cite{ortega2021} introduced two definitions of difference-based multivariate risk contribution measures corresponding to MCoVaR and MCoES, where the benchmark risk measure does not involve systemic risk. However, when regulators in financial markets focus on the relative spillover effects of systemic risk, the effectiveness of these measures becomes limited. To assess relative risk, the relative spillover effect of risk can be measured by dividing the multivariate risk contribution of an entity by its benchmark. One way to evaluate the risk contribution ratio of $X_2, \ldots, X_n$ to $X_1$ is to compare the conditional risk measure of $X_1$ (MCoVaR) with its unconditional risk value (VaR). 
Another method is to replace the unconditional risk value $\mathrm{VaR}$ with the conditional VaR of $X_1$ when $X_2, \ldots, X_n$ are under benchmark conditions, where the benchmark state is typically defined by the median \citep{sordo2018}.

We introduce the definition of a risk contribution ratio measure leveraging MCoVaR as follows. 
\begin{Def}\label{def_RMCoVaR}
For $\bm p = (p_1,\dots,p_n) \in (0,1)^n$, the ratio-based contribution $\rm MCoVaR$ with unconditional $\rm VaR$ as benchmark measure  is defined by\footnote{To avoid misleading, we sometimes use  ${\rm MCoVaR}_{p_1,p_2,\dots,p_n}[X_{1}|X_{2},\dots,X_{n}]$ to represent  ${\rm MCoVaR}_{\bm p}[X_{1}|X_{2},\dots,X_{n}]$.}
\begin{equation}
    \Delta^{\rm R} {\rm MCoVaR}_{\bm p}[X_{1}|X_{2},\dots,X_{n}] =\frac{ {\rm MCoVaR}_{ \bm p}[X_{1}|X_{2},\dots,X_{n}] - {\rm VaR}_{p_1}[X_1]}{{\rm VaR}_{p_1}[X_1]},
\end{equation}
\textcolor{blue}{provided  that \({\rm VaR}_{p_1}[X_1] \neq0\).}
For $p_1 \in (0,1)$, the ratio-based contribution $\rm MCoVaR$ with median-type $\rm MCoVaR$ as the benchmark measure  is defined by
\begin{equation}
    \Delta^{\rm R-med} {\rm MCoVaR}_{\bm p}[X_{1}|X_{2},\dots,X_{n}] =\frac{ {\rm MCoVaR}_{p_1, \bm p_{[-1]}}[X_{1}|X_{2},\dots,X_{n}] - {\rm MCoVaR}_{ p_1, \frac{\bm 1}{\bm 2}}[X_{1}|X_{2},\dots,X_{n}]}{{\rm MCoVaR}_{p_1,\frac{\bm 1}{\bm 2}}[X_{1}|X_{2},\dots,X_{n}]},
\end{equation}
\textcolor{blue}{provided  that \({\rm MCoVaR}_{p_1,\frac{\bm 1}{\bm 2}}[X_{1}|X_{2},\dots,X_{n}] \neq 0\),} where $\bm p_{[-1]} = (p_2,\dots,p_n) \in (1/2,1)^{n-1}$ and $\frac{\bm 1}{\bm 2} = \left(\frac{1}{2},\dots,\frac{1}{2}\right) \in \mathbb R^{n-1}$.
\end{Def}

Similarly, the multivariate risk contribution measures for MCoES are defined as follow.
\begin{Def} \label{def_RMCoES}
For $\bm p = (p_1,\dots,p_n) \in (0,1)^n$, the ratio-based contribution $\rm MCoES$ with unconditional $\rm ES$ as benchmark measure  is defined by
\begin{equation}
    \Delta^{\rm R} {\rm MCoES}_{\bm p}[X_{1}|X_{2},\dots,X_{n}] =\frac{ {\rm MCoES}_{\bm p}[X_{1}|X_{2},\dots,X_{n}] - {\rm ES}_{p_1}[X_1]}{{\rm ES}_{p_1}[X_1]},
\end{equation}
\textcolor{blue}{provided  that \({\rm ES}_{p_1}[X_1] \neq 0\).} For $p_1 \in (0,1)$, the ratio-based contribution $\rm MCoES$ with median-type $\rm MCoES$ as the benchmark measure  is defined by
\begin{equation}
    \Delta^{\rm R-med} {\rm MCoES}_{\bm p}[X_{1}|X_{2},\dots,X_{n}] =\frac{ {\rm MCoES}_{p_1, \bm p_{[-1]}}[X_{1}|X_{2},\dots,X_{n}] - {\rm MCoES}_{ p_1,\frac{\bm 1}{\bm 2}}[X_{1}|X_{2},\dots,X_{n}]}{{\rm MCoES}_{p_1,\frac{\bm 1}{\bm 2}}[X_{1}|X_{2},\dots,X_{n}]},
\end{equation}
\textcolor{blue}{provided  that \({\rm MCoES}_{p_1,\frac{\bm 1}{\bm 2}}[X_{1}|X_{2},\dots,X_{n}] \neq 0\),} where $\bm p_{[-1]} = (p_2,\dots,p_n) \in (1/2,1)^{n-1}$ and $\frac{\bm 1}{\bm 2} = \left(\frac{1}{2},\dots,\frac{1}{2}\right) \in \mathbb R^{n-1}$.
\end{Def}

Correspondingly, the ratio-based contribution $\rm MMME$ with unconditional mean excess as the benchmark measure  is defined as follows.
\begin{Def} \label{def_RMMME}
    For  ${\bm p}_{[-1]} = (p_2,\dots,p_n) \in (0,1)^{n-1}$,  the risk contribution ratio measure of $\rm MMME$ is defined by 
\begin{equation}
\Delta^{\rm R} \mathrm{MMME}_{\bm p_{[-1]} }[X_{1}|X_{2},\dots,X_{n}]= \frac{\mathbb E\left[ \left(X_1-A_{X,p_{[-1]}}\right)_+ \bigg| \bigcap\limits_{j = 2}^n \left\{{X_j} > {\rm VaR}_{p_j}[X_j]\right\} \right] - \mathbb E\left[\left(X_1-A_{X,p_{[-1]}}\right)_+\right]}{\mathbb E[(X_1-A_{X,p_{[-1]}})_+]},
\end{equation}
\textcolor{blue}{provided  that \(\mathbb E[(X_1-A_{X,p_{[-1]}})_+] \neq 0\)}.
\end{Def}


These multivariate risk contribution ratio measures are new compared with the ones introduced in \cite{ortega2021}.  Next, we shall establish sufficient conditions for comparing these newly proposed measures for multivariate risk vectors. For two $n$-dimensional portfolio of risks $\bm X$ and $\bm Y$, this section established sufficient conditions for comparing the ratio-based risk contribution measures in terms of MCoVaR, MCoES, and MMME. 
The next result compares $\Delta^{\rm R} {\rm MCoVaR}_{\bm p}[X_1|X_2,\dots,X_n]$ and $\Delta^{\rm R} {\rm MCoVaR}_{\bm p}[Y_1|Y_2,\dots,Y_n]$ under some appropriate conditions imposed on marginal risks $X_1$ and $Y_1$ and the dependence structure.

\begin{The}\label{The_RCoVaR}
       Let $\bm X = (X_1,\dots,X_n)$ and $\bm Y = (Y_1,\dots,Y_n)$ be two \textcolor{blue}{nonnegative} random vectors with the distribution functions $F$ and $G$, marginal distributions $F_1,\dots,F_n$ and $G_1,\dots,G_n$, and copulas $C$ and $C'$, respectively.  Suppose that $C \leq_{\rm whr} C'$, and either $(X_2,\dots,X_n) \uparrow_{\rm SI} X_1$ or $(Y_2,\dots,Y_n) \uparrow_{\rm SI} Y_1$ holds. Then, for $\bm p=(p_1,\dots,p_n) \in (0,1)^{n}$, $X_1 \leq_{\star} Y_1$ implies that
    \begin{equation} \label{eq_RMCoVaR}
        \Delta^{\rm R} {\rm MCoVaR}_{\bm p}[X_1|X_2,\dots,X_n] \leq \Delta^{\rm R} {\rm MCoVaR}_{\bm p}[Y_1|Y_2,\dots,Y_n].
    \end{equation}

\end{The}

\textcolor{blue}{Recall that an increasing function $h:[0,1]\rightarrow [0,1]$ is said to be a distortion function if it satisfies $h(0)=0$ and $h(1)=1$.} 
The following lemma is needed for proving the comparison results under other types of ratio-based contribution measures.
\begin{Lem} \label{Lem_belzunce}
    \citep{belzunce2012}. Let $X$ and $Y$ be two \textcolor{blue}{nonnegative} random variables with distribution functions $F_X$ and $F_Y$, respectively. Then,
\begin{enumerate} [(i)]
    \item $X \leq_{\star} Y$ if and only if $I_{A,B} (X) \leq I_{A,B}(Y)$ for all distortion function $A(t)$, $B(t)$ and convex function $A \circ B^{-1}(t)$, where
\begin{equation}
    I_{A,B}(X) = \frac{\int_0^1 F_X^{-1}(t) dA(t)}{\int_0^1 F_X^{-1}(t) dB(t)}.
\end{equation}
    \item $X \leq_{\rm ps} Y$ if and only if $I_{A,B}(X) \leq I_{A,B}(Y)$ for all   distortion function $A(t)$, convex distortion function $B(t)$ and convex function $A \circ B^{-1}(t)$.
\end{enumerate}
\end{Lem}

In the following theorem, sufficient conditions for comparison between
 $\Delta^{\rm R} {\rm MCoES}_{\bm p}[X_1|X_2,\dots,X_n]$ and $\Delta^{\rm R} {\rm MCoES}_{\bm p}[Y_1|Y_2,\dots,Y_n]$ are provided in terms of the expected proportional shortfall order.

\begin{The} \label{The_RCoES}
Let $\bm X = (X_1,\dots,X_n)$ and $\bm Y = (Y_1,\dots,Y_n)$ be two \textcolor{blue}{nonnegative}  random vectors with distribution functions $F$ and $G$, marginal distributions $F_1,\dots,F_n$ and $G_1,\dots,G_n$, and copulas $C$ and $C'$, respectively.  Suppose that $C \leq_{\rm whr} C'$, and either $(X_2,\dots,X_n) \uparrow_{\rm SI} X_1$ or $(Y_2,\dots,Y_n) \uparrow_{\rm SI} Y_1$ holds. Then, for $\bm p=(p_1,\dots,p_n) \in (0,1)^{n}$, $X_1 \leq_{\rm ps} Y_1$ implies that
    \begin{equation} \label{eq_RMCoES}
        \Delta^{\rm R} {\rm MCoES}_{\bm p}[X_1|X_2,\dots,X_n] \leq \Delta^{\rm R} {\rm MCoES}_{\bm p}[Y_1|Y_2,\dots,Y_n].
    \end{equation}
\end{The}

Next, we provide an example to illustrate the findings in Theorems \ref{The_RCoVaR} and \ref{The_RCoES}. It is known that if a random vector $\bm X$ satisfies ${\rm MTP}_2$, it then implies that $\hat{\bm X}_i \uparrow_{\rm SI} X_i$ holds for all $i \in \{1, \dots, n\}$. An example of a copula satisfying ${\rm MTP}_2$ is provided below.
As per \citet{muller2005}, the Archimedean copula $C_\psi$ is ${\rm MTP}_2$ if and only if $(-1)^n\psi^{(n)}$ is log-convex,  where $\psi^{(n)}$ denotes the $n$-th derivative. Let $\Psi^{(n)}:= {\rm ln} \left( (-1)^n \psi^{(n)} \right)$. For the Gumbel copula, $\psi (u) = (-{\rm ln}~u)^\theta$, with $n=3$, we have
\begin{equation*}
    C_\theta(u_1,u_2,u_3) = {\rm exp} \left\{ -\left[(-\rm ln~u_1)^\theta + (-\rm ln~u_2)^\theta +(-\rm ln~u_3)^\theta \right]^{\frac{1}{\theta}}\right\}.
\end{equation*}
Fixing  $\theta=2$, it follows that
\begin{equation*}
    \Psi^{(3)}(u) = {\rm ln} \left( \frac{6-4 \cdot {\rm ln}~u}{u^3} \right).
\end{equation*}
Furthermore,  taking the second derivative with respect to $u$ yields that
\begin{equation*}
    \frac{d^2 \Psi^{(3)}(u)}{du^2} = \frac{12({\rm ln}~u)^2 - 40 \cdot {\rm ln}~u +29}{u^2(2 \cdot {\rm ln}~u-3)^2}>0,~ u \in (0,1),
\end{equation*}
which implies that $C_2(u_1,u_2,u_3)$ is ${\rm MTP}_2$. Based on this observation, the following two examples can be established.

\begin{figure}[htbp]
    \centering
    \includegraphics[width=15cm,height=8cm]{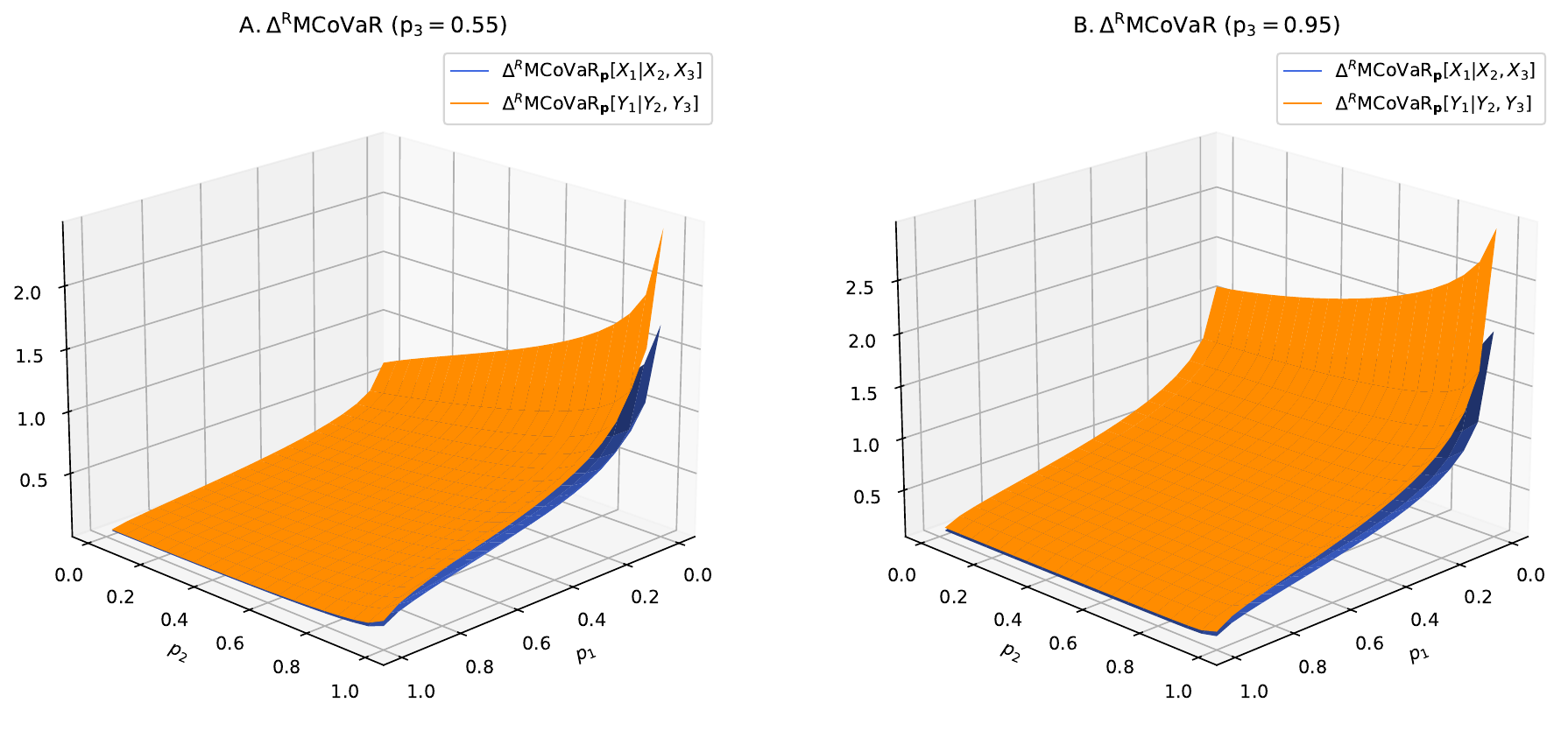}
\caption{Plots of $\Delta^{\rm R} {\rm MCoVaR}_{\bm p}[X_1|X_2,X_3]$ and $\Delta^{\rm R} {\rm MCoVaR}_{\bm p}[Y_1|Y_2,Y_3]$.}
    \label{fig_RMCoVaR}
\end{figure}

\begin{Exa}
    Let $\bm X = (X_1,X_2,X_3)$ and $\bm Y=(Y_1,Y_2,Y_3)$ are two random vectors with Gumbel copula $C$ and $C'$, respectively. By taking $\theta = 2$, it satisfies $(X_2,X_3) \uparrow_{\rm SI} X_1$.  We denote by $Z\sim W(\alpha,\beta)$ to state that $Z$ has a Weibull distribution with scale parameter $\alpha>0$ and shape parameter $\beta>0$. Suppose $X_1 \sim W(1,5)$ and $Y_1 \sim W(1,4)$, indicating $X_1 \leq_{\star} Y_1$ \textcolor{blue}{(see Table 2.1 on p.102 of \citet{belzunce2015introduction})}. As plotted in Figure \ref{fig_RMCoVaR}, the result of Theorem \ref{The_RCoVaR} is illustrated.
\end{Exa}

\begin{figure}[htbp]
    \centering
    \includegraphics[width=15cm,height=8cm]{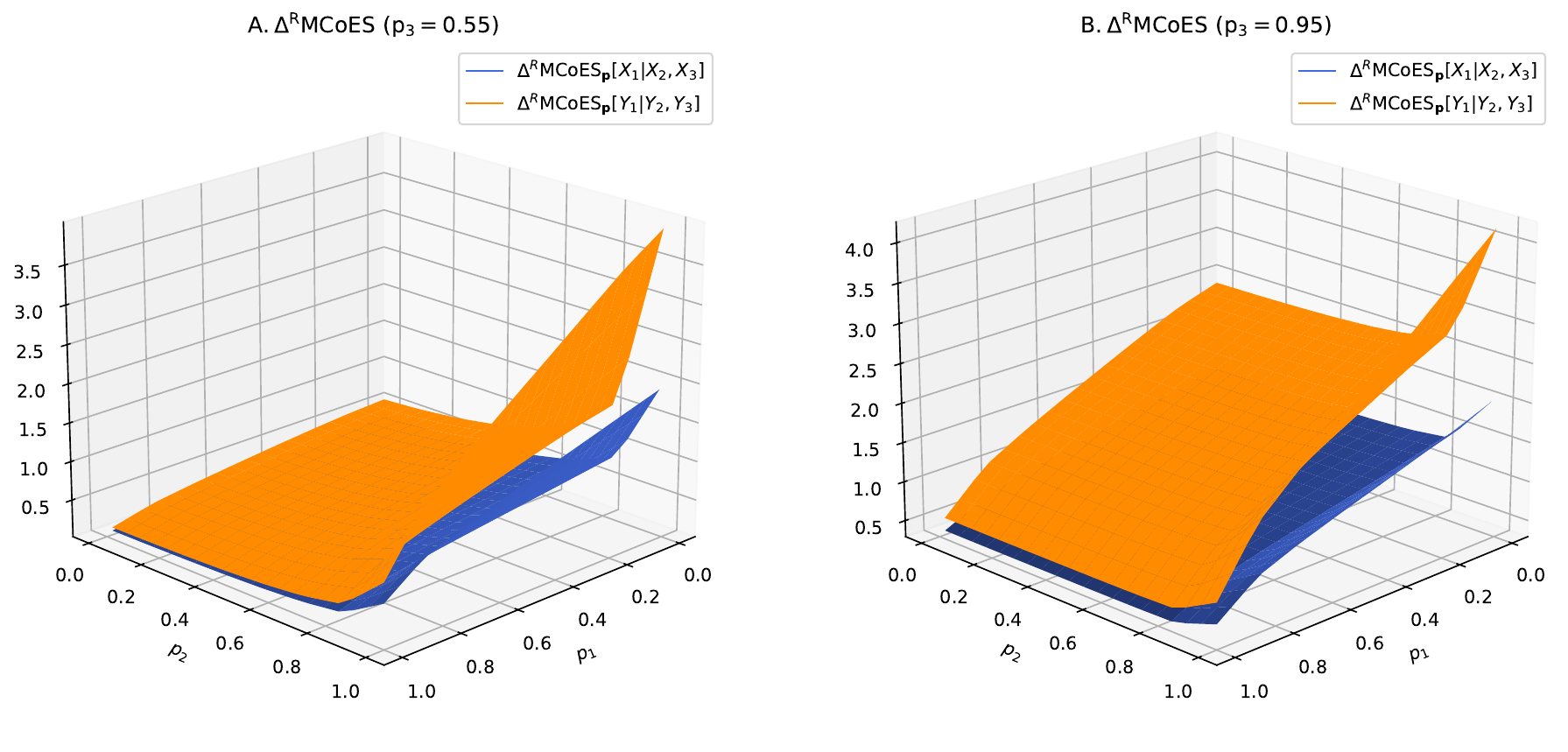}
\caption{Plots of $\Delta^{\rm R} {\rm MCoES}_{\bm p}[X_1|X_2,X_3]$ and $\Delta^{\rm R} {\rm MCoES}_{\bm p}[Y_1|Y_2,Y_3]$.}
    \label{fig_RMCoES}
\end{figure}

\begin{Exa}
    Let $\bm X = (X_1,X_2,X_3)$ and $\bm Y=(Y_1,Y_2,Y_3)$ are two random vectors with Gumbel copula $C$ and $C'$, respectively. By taking $\theta = 2$, it satisfies $(X_2,X_3) \uparrow_{\rm SI} X_1$.  
We denote $Z \sim G(\alpha,\beta)$ to represent that the random variable $Z$ follows the Gamma distribution with shape parameter $\alpha>0$ and scale parameter $\beta>0$. 
\textcolor{blue}{Suppose $X_1 \sim G(3,1)$ and $Y_1 \sim G(1,1)$}, indicating $X_1 \leq_{\star} Y_1$ \textcolor{blue}{(see Table 2.1 on p.102 of \citet{belzunce2015introduction})}. As plotted in Figure \ref{fig_RMCoES}, the result of Theorem \ref{The_RCoES} is illustrated.
\end{Exa}

The next result compares $\Delta^{\rm R} \mathrm{MMME}_{\bm p_{[-1]} }[X_{1}|X_{2},\dots,X_{n}]$ and $\Delta^{\rm R} \mathrm{MMME}_{\bm p_{[-1]} }[Y_{1}|Y_{2},\dots,Y_{n}]$ under some appropriate conditions imposed on the dependence structure when $\bm X$ and $\bm Y$ have the same marginals. 

\begin{The}
 \label{The_RMMMES}
 Let $\bm X = (X_1,\dots,X_n)$ and $\bm Y = (Y_1,\dots,Y_n)$ be two random vectors with the distribution functions $F$ and $G$ and same marginal distributions.  If $C \leq_{\rm whr} C'$, then for all $\bm p_{[-1]} = (p_2,\dots,p_n) \in (0,1)^{n-1}$, we have
    \begin{equation} \label{eq_RMMME}
        \Delta^{\rm R} \mathrm{MMME}_{\bm p_{[-1]} }[X_{1}|X_{2},\dots,X_{n}] \leq \Delta^{\rm R} \mathrm{MMME}_{\bm p_{[-1]} }[Y_{1}|Y_{2},\dots,Y_{n}].
    \end{equation}
\end{The}

In Theorem \ref{The_RMMMES}, we assumed that both vectors have the same marginal distributions. Under this assumption, the comparison of $\Delta^{\rm R}{\rm MMME}$ can be equivalent to the comparison of MMME, with conditions similar to Corollary 2 in \citet{ortega2021}. Therefore, the two measures exhibit consistent ordering under the copula structure, a property we refer to as dependence consistency. 

\begin{figure}[htp]
    \centering
    \includegraphics[width=8cm,height=8cm]{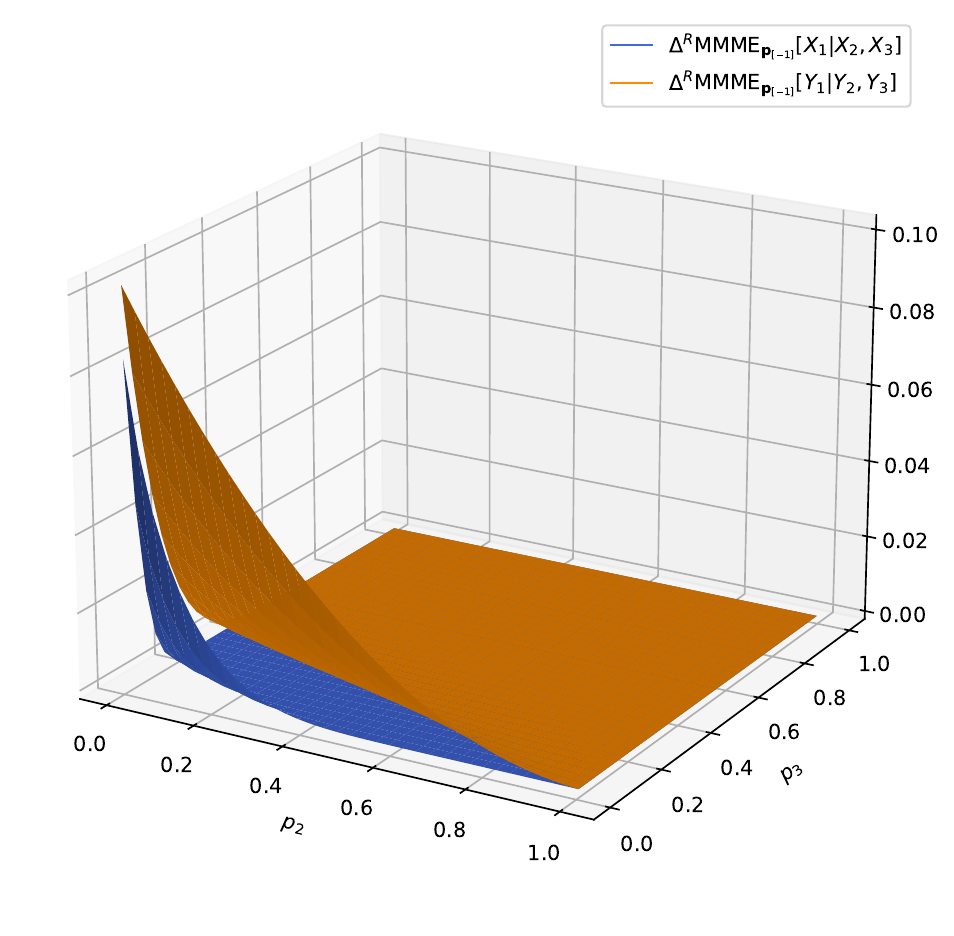}
\caption{Plots of $\Delta^{\rm R} {\rm MMME}_{\bm p_{[-1]}}[X_1|X_2,X_3]$ and $\Delta^{\rm R} {\rm MMME}_{\bm p_{[-1]}}[Y_1|Y_2,Y_3]$.}
    \label{fig_RMMME}
\end{figure}

The following example is provided to show the validity of Theorem \ref{The_RMMMES}.
\begin{Exa}
  Let $\bm X$ be an $n$-dimensional random vector following the multivariate Gumbel Exponential distribution with its joint survival function given by
  \begin{equation*}
      \overline F_{\bm \lambda}(x_1,\dots,x_n) = {\rm exp} \left\{ -\sum_I \lambda_I \prod_{i \in I} x_i \right\},~x_i \geq 0,~i=1,\dots,n,
  \end{equation*}
  where $\bm \lambda = \{\lambda_I: I  \subseteq \{1,\dots,n\},\lambda_I \geq 0, I \ne  \emptyset \}$.  Let $\bm Y$ be another $n$-dimensional random vector with a multivariate Gumbel Exponential survival distribution $\overline G_{{\bm \lambda}^*}$. For \( n=3 \), let \( \lambda^*_i =\lambda_i = 10 \), which implies \( X_i \stackrel{\rm st}{=} Y_i \) for \( i=1,2,3 \). Additionally, set \( \lambda^*_{12} = \lambda^*_{13} = \lambda^*_{23} = \lambda^*_{123} = 10 \) and \( \lambda_{12} = \lambda_{13} = \lambda_{23} = \lambda_{123} = 100 \). Since \( \bm{\lambda} \geq \bm{\lambda}^* \), according to \citet{khaledi2005}, this implies \( \bm{X} \leq_{\rm whr} \bm{Y} \) (or \( C \leq_{\rm whr} C' \)). The plots of $\Delta^{\rm R} {\rm MMME}_{\bm p_{[-1]}}[X_1|X_2,X_3]$ and $\Delta^{\rm R} {\rm MMME}_{\bm p_{[-1]}}[Y_1|Y_2,Y_3]$ are displayed in Figure \ref{fig_RMMME}, which is consistent with the result of Theorem \ref{The_RMMMES}.

\end{Exa}

Under an unconditional risk measure as benchmark, different copulas can be used to measure the risk contributions of two multivariate risk vectors, as the benchmark is unaffected by the copula. This is the reason why we can consider different copulas in the comparison results developed in Theorems \ref{The_RCoVaR}, \ref{The_RCoES} and \ref{The_RMMMES}. However, when a median-type co-risk measure as the benchmark, which is also affected by the copula, using different copulas can make the comparison of risk contribution measures challenging. Therefore, to ensure the comparability of median-type risk contribution measures, the same copula will be adopted in the following discussions.


The next result establishes sufficient conditions for comparison between
 $\Delta^{\rm R-med} {\rm MCoVaR}_{\bm p}[X_1|X_2,\dots,X_n]$ and $\Delta^{\rm R-med} {\rm MCoVaR}_{\bm p}[Y_1|Y_2,\dots,Y_n]$ when $X_1$ and $Y_1$ are ranked by the star order.

\begin{The}
\label{The_RmedCoVaR}
       Let $\bm X = (X_1,\dots,X_n)$ and $\bm Y = (Y_1,\dots,Y_n)$ be two \textcolor{blue}{nonnegative} random vectors with  distribution functions $F$ and $G$, and marginal distributions are $F_1,\dots,F_n$ and $G_1,\dots,G_n$, respectively. Suppose that $\bm X$ and $\bm Y$ have the same copula $C$ such that $X_1 \uparrow_{\rm RTI} (X_2,\dots,X_n)$. For $p_1 \in (0,1)$ and $(p_2,\dots,p_n) \in (1/2,1)^{n-1}$, $X_1 \leq_{\star} Y_1$ implies that
    \begin{equation} \label{eq_RmedMCoVaR}
        \Delta^{\rm R-med} {\rm MCoVaR}_{\bm p}[X_1|X_2,\dots,X_n] \leq \Delta^{\rm R-med} {\rm MCoVaR}_{\bm p}[Y_1|Y_2,\dots,Y_n].
    \end{equation}
\end{The}

It is worth addressing that the condition \(X_1 \uparrow_{\rm RTI} (X_2, \dots, X_n)\) indicates that the conditional probability \(\mathbb{P}(X_1 > x_1 \mid X_2 > x_2, \dots, X_n > x_n)\) increases with \(x_j\), focusing on how \(X_1\) behaves given \((X_2, \dots, X_n)\). Such condition can be implied from \(X_1 \uparrow_{\rm SI} (X_2, \dots, X_n)\), which is very different from \((X_2, \dots, X_n) \uparrow_{\rm SI} X_1\). In fact, the later condition \((X_2, \dots, X_n) \uparrow_{\rm SI} X_1\) means that the conditional distribution of  $(X_2,\dots,X_n|X_1=x_1)$ is stochastically increasing in \(x_1\), showing how the random vector \((X_2, \dots, X_n)\) shifts as \(X_1\) changes. These conditions describe different stochastic relationships and do not imply any interdeducible relationship.


In the following theorem, we establish sufficient conditions for comparing 
 $\Delta^{\rm R-med} {\rm MCoES}_{\bm p}[X_1|X_2,\dots,X_n]$ and $\Delta^{\rm R-med} {\rm MCoES}_{\bm p}[Y_1|Y_2,\dots,Y_n]$ when $X_1$ and $Y_1$ are ranked by the expected proportional shortfall order.

\begin{The}\label{The_RmedCoES}
 Let $\bm X = (X_1,\dots,X_n)$ and $\bm Y = (Y_1,\dots,Y_n)$ be two \textcolor{blue}{nonnegative} random vectors with  distribution functions $F$ and $G$, and marginal distributions are $F_1,\dots,F_n$ and $G_1,\dots,G_n$, respectively. Suppose that $\bm X$ and $\bm Y$ have the same copula $C$ such that $C$ is ${\rm MTP}_2$. For $p_1 \in (0,1)$ and $(p_2,\dots,p_n) \in (1/2,1)^{n-1}$, $X_1 \leq_{\rm ps} Y_1$ implies that
    \begin{equation} \label{eq_RmedMCoES}
        \Delta^{\rm R-med} {\rm MCoES}_{\bm p}[X_1|X_2,\dots,X_n] \leq \Delta^{\rm R-med} {\rm MCoES}_{\bm p}[Y_1|Y_2,\dots,Y_n].
    \end{equation}
\end{The}

 \begin{figure}[htbp]
    \centering
    \includegraphics[width=15cm,height=8cm]{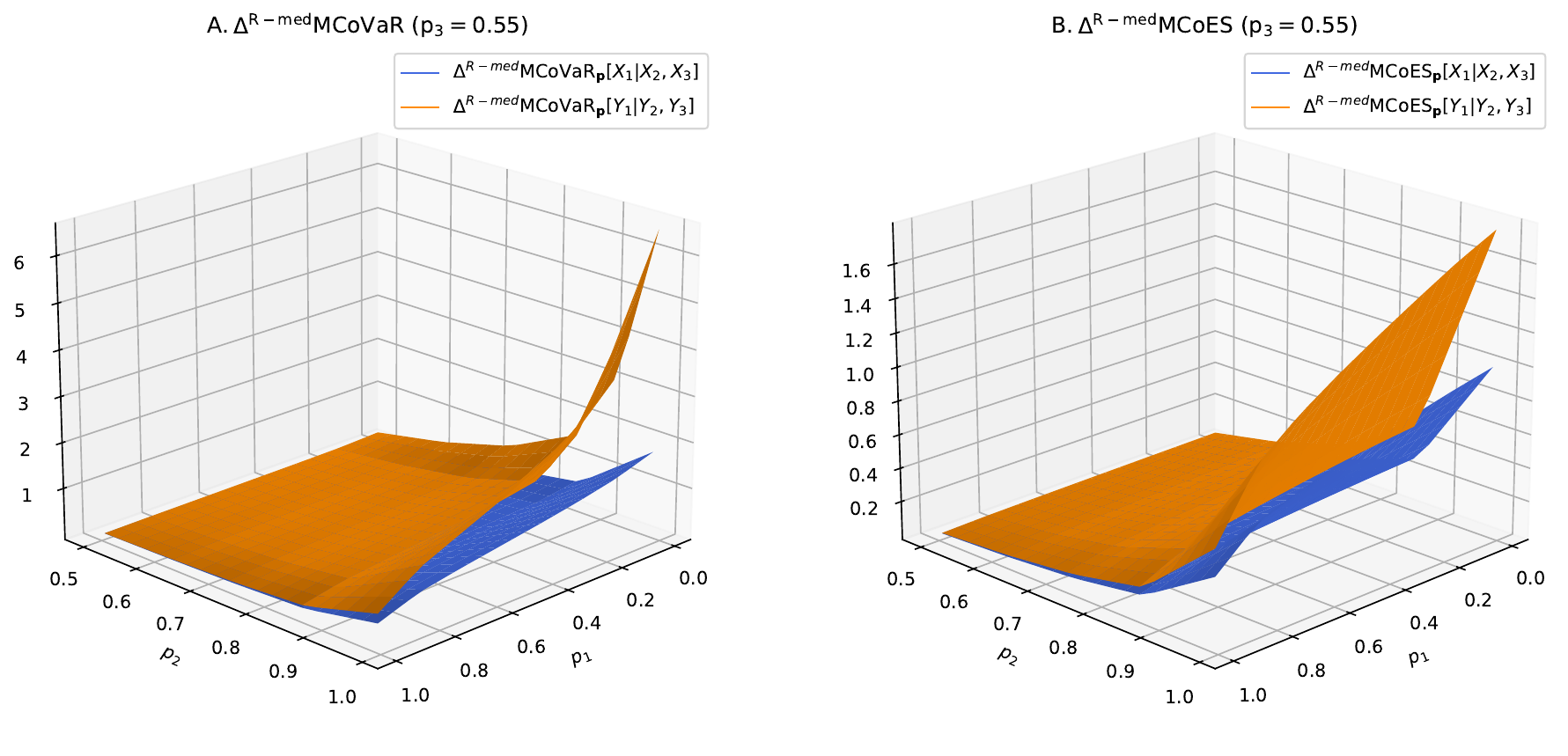}
\caption{Subfigure A: plots of $\Delta^{\rm R-med} {\rm MCoVaR}_{\bm p}[X_1|X_2,X_3]$ and $\Delta^{\rm R-med} {\rm MCoVaR}_{\bm p}[Y_1|Y_2,Y_3]$. Subfigure B: plots of $\Delta^{\rm R-med} {\rm MCoES}_{\bm p}[X_1|X_2,X_3]$ and $\Delta^{\rm R-med} {\rm MCoES}_{\bm p}[Y_1|Y_2,Y_3]$.}\label{fig_RmedMCo}
\end{figure}

The following example illustrates the results of Theorems \ref{The_RmedCoVaR} and \ref{The_RmedCoES}.
\begin{Exa}
    Let \( \bm{X} = (X_1,X_2,X_3) \) and \( \bm{Y} = (Y_1,Y_2,Y_3) \) be two random vectors with the same Gumbel copula \( C \) with \( \theta = 2 \). Thus, \( C \) is ${\rm MTP}_2$ and \( X_1 \uparrow_{\rm RTI} (X_2, X_3) \). Suppose $X_1 \sim G(3,1)$ and $Y_1 \sim G(1,1)$, which implies $X_1 \leq_{\star{\rm [ps]}} Y_1$ \textcolor{blue}{(see Table 2.2 on p.102 of \citet{belzunce2015introduction}). 
The plots of $\Delta^{\rm R-med} {\rm MCoVaR}$  and  $\Delta^{\rm R-med} {\rm MCoES}$  are shown in Figure \ref{fig_RmedMCo}, which agree with both of the results in Theorem \ref{The_RmedCoVaR} and Theorem \ref{The_RmedCoES}.}
\end{Exa}


\section{An application in the cryptocurrency market}\label{sec:Empirical}
In this section, risk measures proposed in Section \ref{sec:mulcontri} are applied for cryptocurrency market dataset to quantify the relative spillover effects. \textcolor{blue}{We  employ a static methodology to quantify the relative risk contributions of cryptocurrencies using the multivariate risk measures discussed in this paper. The static approach offers several distinct advantages that align with the goals of this study. Firstly, it delivers clear and interpretable results, allowing for a straightforward comparison of risk contributions among different assets. This clarity is essential for both theoretical validation and practical application, as it provides actionable insights for investors and regulators. Secondly, the static method is computationally efficient, enabling the analysis of large datasets and the application of complex risk measures without excessive resource demands. This efficiency is particularly valuable in the context of cryptocurrencies, where data volumes are substantial and market dynamics are intricate. Lastly, the static approach provides a stable framework for initial exploration and validation of new risk measurement tools, ensuring that the core properties of these tools can be assessed reliably before extending to more complex dynamic analyses}.
\par The analysis utilizes three cryptocurrencies (CCs): Bitcoin (BTC), Ethereum (ETH), and Monero (XMR). The data contains daily closing prices in USD stemming from the Community Network Data provided by CoinMetrics.\footnote{See \url{https://coinmetrics.io/}.} The sample includes $N = 3226$ observations from 01/09/2015 to 30/06/2024 as CCs are traded every day, including weekends. For ease of our subsequent analysis, the prices are transformed in log-losses, that is,
\begin{equation}\nonumber
    X_{i,t}=-100 \cdot \ln\left(p_{i,t} / p_{i,t-1}\right),
\end{equation}
where $X_{i,t}$ represents the percentage-based log-losses\footnote{Differing from the majority of economics and finance literature, this paper focuses on risk measurement based on the probability distributions of losses. Since we denote {$L_t$} as the stock index's declining changing pattern, the right tail of the distribution of {$L_t$} represents extreme risk.} on day $t$  with $i \in \{1, 2, 3\}$ for BTC, ETH, and XRP and $p_{i,t}$ denotes the closing price for cryptocurrency $i$ on day $t$. For each cryptocurrency $i$, $N$ observations $(x_{i,1}, \dots, x_{i,N})$ are obtained. A statistical summary for these percentage-based log-loss samples is shown in Table \ref{tab:summary}, and their Spearman and Kendall correlation matrices are provided in Table \ref{tab:corr}. From Table \ref{tab:summary}, it can be observed that both BTC and XMR have relatively low mean losses, with BTC showing the smallest standard deviation, indicating a more stable performance. 
\textcolor{blue}{The elevated standard deviations observed for Ethereum and Monero relative to Bitcoin may reflect heightened exposure to volatility inherent in the cryptocurrency market, particularly during discrete events or periods of acute market stress. Ethereum, for instance, has demonstrated sensitivity to regulatory scrutiny and protocol upgrades (e.g., transitions to Ethereum 2.0), while Monero's volatility has been amplified by debates over privacy regulations and network congestion. These idiosyncratic factors, coupled with broader market uncertainty, likely contribute to the pronounced fluctuations seen in both assets compared to Bitcoin.} 
The results from Table \ref{tab:corr} indicate that all the three cryptocurrencies enjoy positive dependence structure in losses, and the correlation between BTC and ETH is relatively stronger compared with the correlation between ETH and XMR.



\begin{table}[ht]
\centering
\caption{Statistical summary for log-losses of the cryptocurrencies}
\label{tab:summary}
\begin{tabular}{cccccc}
\hline
Cryptocurrency & Mean   & Median & Max   & Min    & Standard Deviation  \\ \hline
BTC              & -0.174 & -0.171 & 47.056 & -22.405 & 3.705 \\
ETH              & -0.243 & -0.054 & 56.561 & -30.062 & 5.527 \\
XMR              & -0.183 & -0.203 & 49.224 & -59.634 & 5.625 \\ \hline
\end{tabular}
\end{table}

\begin{table}[tph]
 \renewcommand{\arraystretch}{1.1}
\caption{Correlation matrix for log-losses of cryptocurrencies}
 \label{tab:corr}
\centering
\label{table1}
\begin{tabular}{p{2cm}p{2cm}p{2cm}p{2cm}}
\hline
\multicolumn{4}{c}{Panel A: Spearman correlation matrix} \\ \hline
           & BTC   & ETH   &  XMR   \\ \cline{2-4}
BTC        & 1.000 & 0.630 & 0.575 \\
ETH        & 0.630 & 1.000 & 0.565 \\
XMR        & 0.575 & 0.565 & 1.000 \\ \hline
\multicolumn{4}{c}{Panel B: Kendall correlation matrix} \\ \hline
           & BTC   & ETH   &  XMR   \\  \cline{2-4}
BTC        & 1.000 & 0.486 & 0.424 \\
ETH        & 0.486 & 1.000 & 0.418 \\
XMR        & 0.424 & 0.418 & 1.000 \\ \hline
\end{tabular}
\end{table}

  
\par We use the method of inference function for margins (IFM) proposed in \cite{joe2005asymptotic}, which has been also widely applied by many research papers such as \cite{shi2018pair} and \cite{zhu2023asymptotic}. Interested readers can refer to \cite{joe2005asymptotic} for more efficiency properties of IFM. To calculate the co-risk measures introduced both in Sections \ref{sec:pre} and \ref{sec:mulcontri}, the following empirical procedure is adopted:
\begin{itemize}
    \item \underline{Step 1}: Estimate the respective marginal model for each group of samples separately. 
    \item \underline{Step 2}: Estimate the copula based on the pseudo-sample observations obtained from the parametric probability integral transformation on the samples.
    \item \underline{Step 3}: Calculate the respective risk measures.
\end{itemize}

\subsection{Parameter estimation of the marginal distributions}
The Generalized Pareto Distribution (GPD) is widely applied in the insurance and finance sectors as a probability distribution for modeling extreme values \citep{embrechts2013modelling,castillo1997fitting}. Its distribution function is defined as follows:
\begin{equation} \label{eq_marginal F}
F(x;\xi ,\beta ) = \left\{ \begin{array}{l}
1 - \left(1+\frac{\xi}{\beta}x \right)^{-\frac{1}{\xi}},~~\xi \ne 0\\
1 - {\rm exp}\left(-\frac{x}{\beta} \right),~~~~~\xi =0
\end{array} \right.,
\end{equation}
where $\xi$ is the shape parameter and $\beta$ is the scale parameter. For each individual risk $X_i$ with distribution function $F_i$ and a confidential level $\alpha$, the GPD serves as a suitable approximation for the excess distribution function $\mathbb P(X_i - \alpha \leq x| X_i > \alpha)$. Consequently, the following approximation can be utilized:
\begin{eqnarray*}
    F_i(x) &=& F_i(\alpha) +\mathbb P(X_i-\alpha \leq x-\alpha| X_i>\alpha)\left( 1 - F_i(\alpha) \right) \\
    &\approx& F_i(\alpha) + F(x-\alpha;\xi_i,\beta_i)\left( 1 - F_i(\alpha) \right),~ x\geq \alpha.
\end{eqnarray*}

For $t=1,\dots,T$, let $x_{i,(1)} \leq\dots \leq x_{i,(T)}$ denote the order statistics of $\{x_{i,t} \}_{1 \leq t \leq T}$, $i=1,2,3$. Setting threshold $\alpha$, we apply the empirical distribution to fit the samples that are less than the $\alpha$-quantile and utilize the GPD to fit the samples that are greater than the $\alpha$-quantile. Using the Maximum Likelihood Estimation (MLE) method to estimate parameters $\xi_i$ and $\beta_i$, denote as $\hat{\xi}_i$ and $\hat{\beta}_i$, the estimated marginal distribution $F_i$ is as follows:
\begin{equation}\label{eq:Fhatxin0}
\hat{F}_i(x) = \left\{
    \begin{array}{l}
\frac{\sum_{t=1}^T {\rm {\bm I}}(x_{i,t} \leq x)}{T},~~~~~~~~~~~~~~~~~~~~~~~~ x \leq x_{i,(\left\lceil \alpha T \right\rceil)} \\
        1 - (1-\alpha) \exp\left( -\frac{x - x_{i,(\left\lceil \alpha T \right\rceil)}}{\hat{\beta}_i} \right), ~~ x > x_{i,(\left\lceil \alpha T \right\rceil)}
    \end{array}
\right.
\end{equation}
when $\hat \xi_i = 0$,\footnote{\(\left\lceil k \right\rceil\) denotes the ceiling of \( k \), i.e., rounding \( k \) up to the nearest integer.} and
\begin{equation} \label{eq:Fhatxi0}
    \hat{F}_i(x) = \left\{ \begin{array}{l}
        \frac{\sum_{t=1}^T {\rm {\bm I}}(x_{i,t} \leq x)}{T}, ~~~~~~~~~~~~~~~~~~~~~~~~~~~~~~x \leq x_{i,(\left\lceil \alpha T \right\rceil)} \\
         1-(1-\alpha)\left( 1 + \frac{\hat{\xi}_i}{\hat{\beta}_i} (x -x_{i,(\left\lceil \alpha T \right\rceil)})\right),~~x > x_{i,(\left\lceil \alpha T \right\rceil)}
    \end{array}
\right.
\end{equation}
when $\hat \xi_i \ne 0$.

\begin{figure}[htbp!]
        \centering
        \subfigure[QQ plot for BTC based on GPD.]{
        \includegraphics[width=4.5cm]{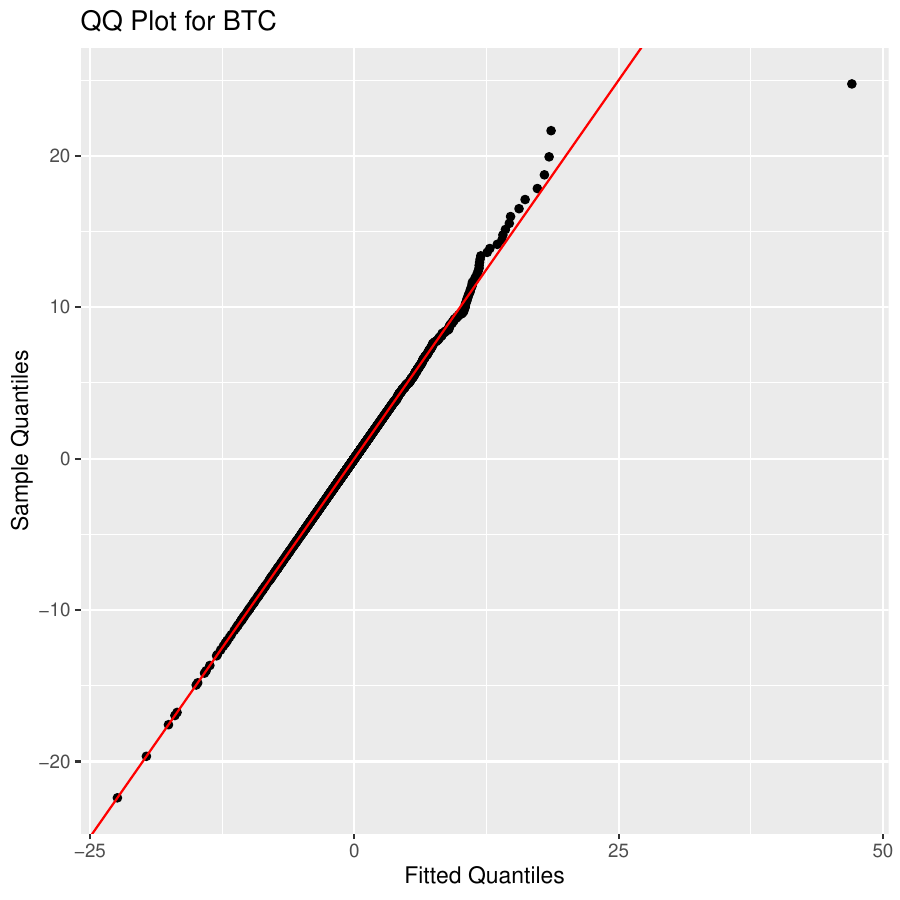}
        }
        \quad
        \subfigure[QQ plot for ETH based on GPD.]{
        \includegraphics[width=4.5cm]{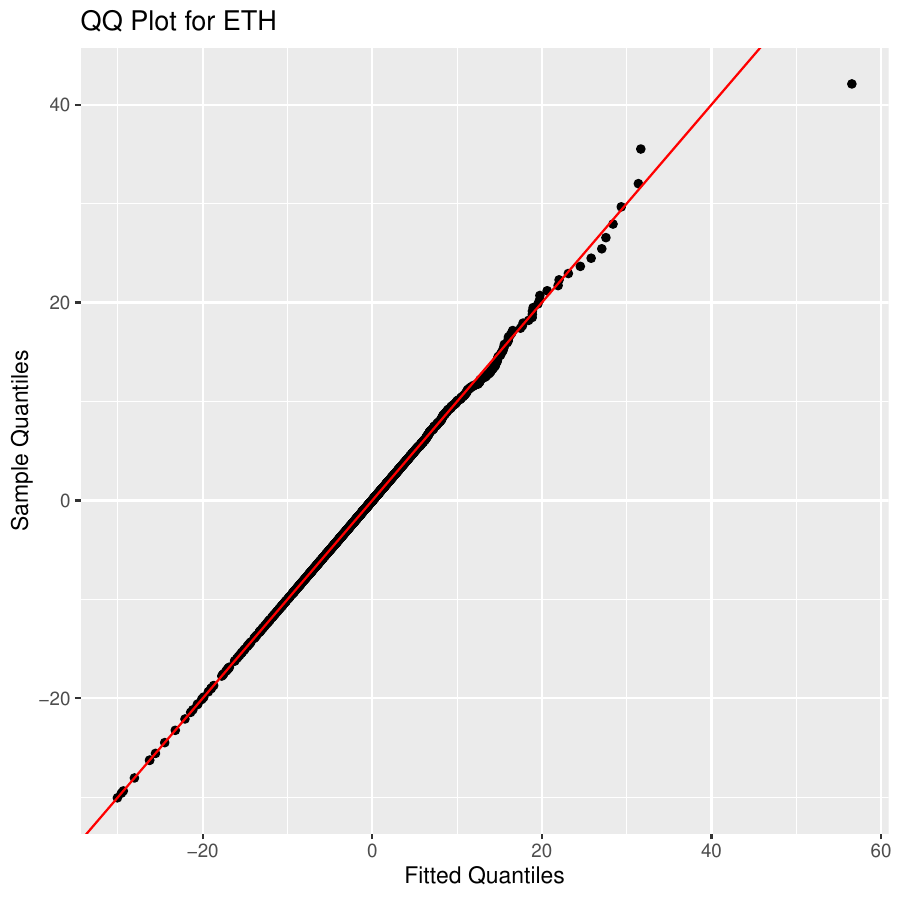}
        }
        \quad
        \subfigure[QQ plot for XMR based on GPD.]{
        \includegraphics[width=4.5cm]{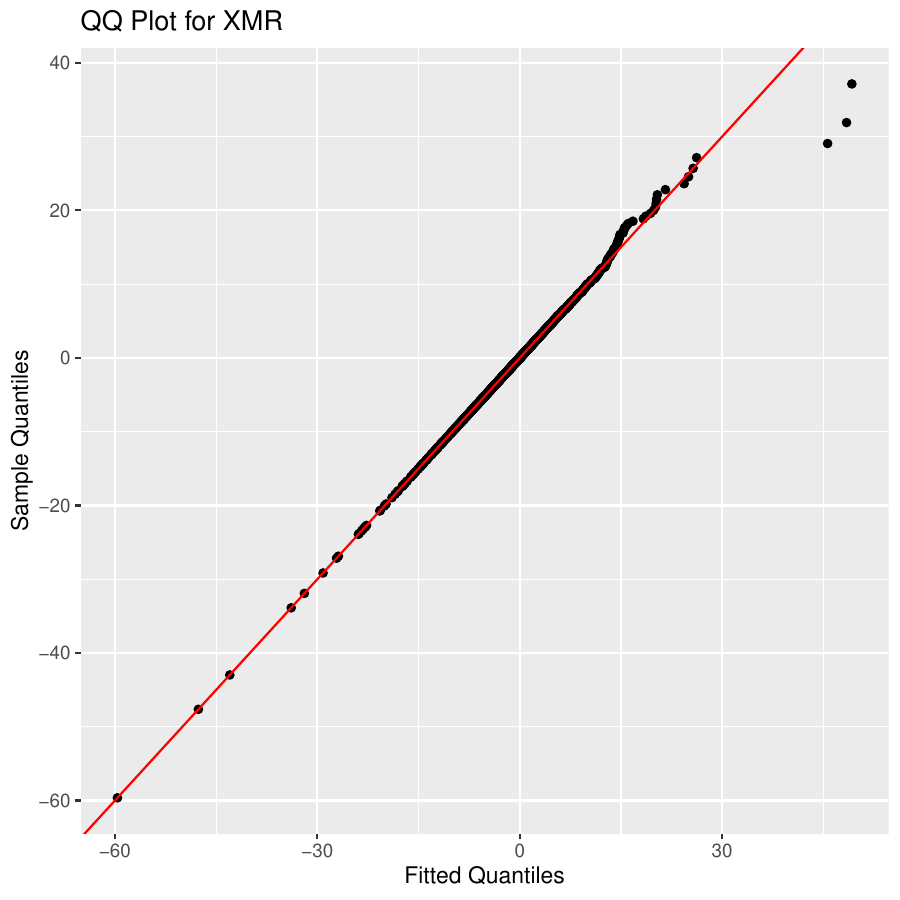}
        }
        \quad
        \subfigure[QQ plot for BTC based on normal distribution.]{
        \includegraphics[width=4.5cm]{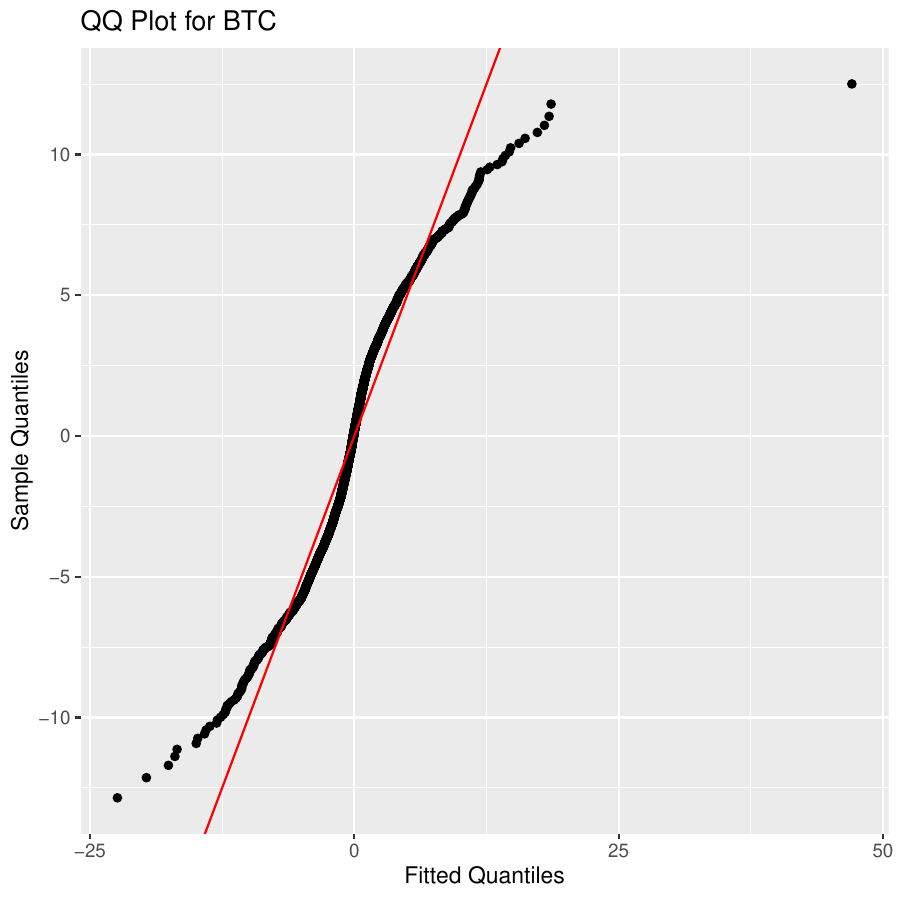}
        }
        \quad
        \subfigure[QQ plot for ETH based on normal distribution.]{
        \includegraphics[width=4.5cm]{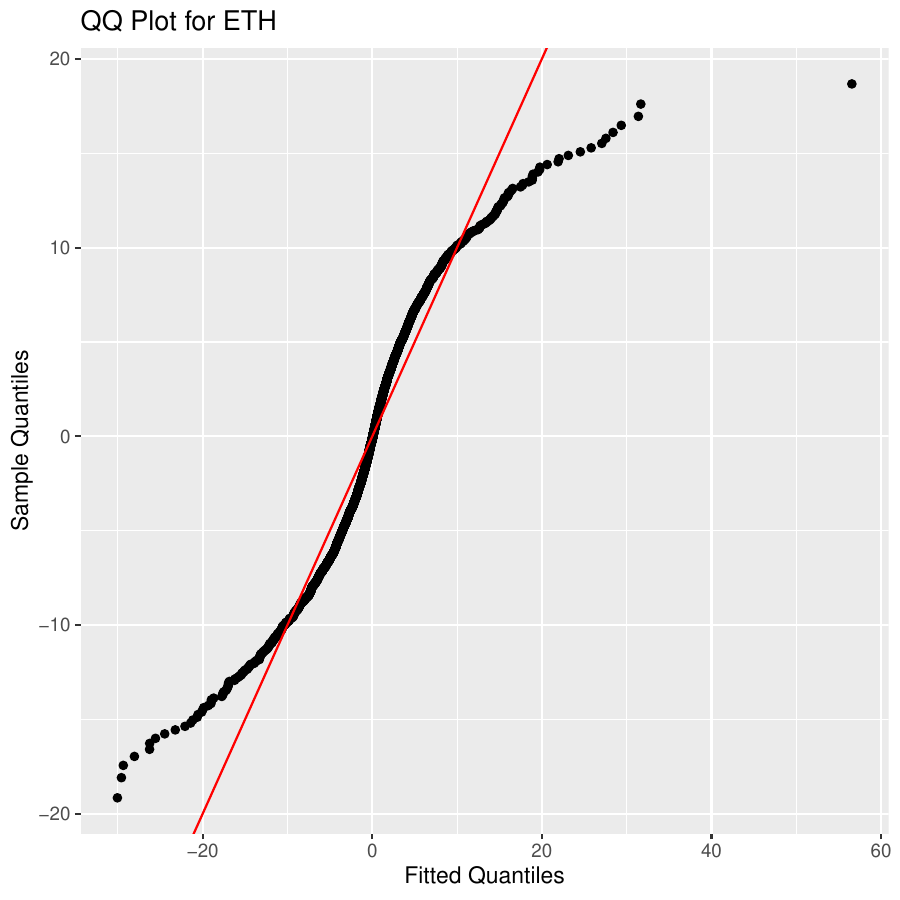}
        }
        \quad
        \subfigure[QQ plot for XMR based on normal distribution.]{
        \includegraphics[width=4.5cm]{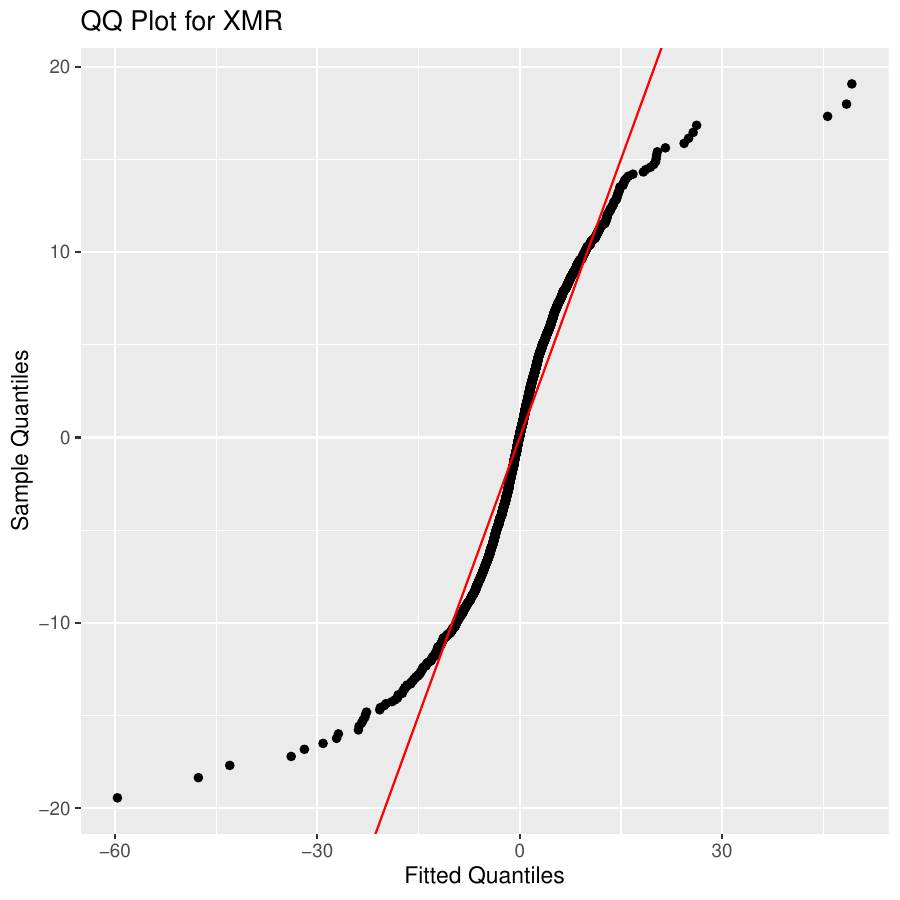}
        }
        \caption{QQ plots for all three cryptocurrencies.}
        \label{fig:GPDQQplot}
        \end{figure}

We set the threshold $\alpha = 90\%$ \citep[cf.][]{koliai2016extreme}, and then use MLE to estimate parameters $\xi_i$ and $\beta_i$, $i = 1, 2, 3$ of GPDs. MLE estimations $\hat{\xi_i}$ and $\hat{\beta_i}$ of $\xi_i$ and $\beta_i$ for three cryptocurrencies are shown in Table \ref{tab:mleresult}. The QQ plots based on the GPD and the normal distribution for all three cryptocurrencies are provided in Figure \ref{fig:GPDQQplot}, which illustrates that the GPD provides significantly better fits for the tails of the data than the normal distribution.

\begin{table}[ht]
    \caption{MLE results for the three cryptocurrencies}
    \centering
    \label{tab:mleresult}
    \begin{tabular}{cccc}
    \hline
    CC                   & Parameter               & Estimation  & Std. Error  \\ \hline
    \multirow{2}{*}{BTC} & $\xi_1$   & 0.083  & 0.051  \\
                         & $\beta_1$ & 2.841  & 0.214 \\ \hline
    \multirow{2}{*}{ETH} & $\xi_2$   & 0.171  & 0.067  \\
                         & $\beta_2$ & 3.757 & 0.325 \\ \hline
    \multirow{2}{*}{XMR} & $\xi_3$   & 0.136  & 0.056  \\
                         & $\beta_3$ & 3.612 & 0.283 \\ \hline
    \end{tabular}
    \end{table}
\par It must be pointed out that GPD degenerates to Exponential distribution when $\xi =0$, according to \eqref{eq_marginal F}. Hence, the expressions of estimated marginal distributions depend on whether the values of $\xi_i$ are nonzero or not, for $i = 1, 2, 3$. Based on the method developed in \cite{tsay2014introduction},
we carry on testing $\xi_i \neq$ 0 by applying the principle that 0 is not in the 95\% confidence interval
[$\hat{\xi_i} - 1.96 \times$ SE($\hat{\xi_i}$), $\hat{\xi_i} + 1.96 \times $SE($\hat{\xi_i}$)] of the estimation $\xi_i$, where SE($\hat{\xi_i}$) denotes the standard error of $\hat{\xi_i}$.\footnote{The test can be also implemented for the case $\xi_1 = 0$. The results, which can be provided upon request, show that the computed values of various risk measures are very close, and thus we do not show them repeatedly here.} As shown in Table \ref{tab:mleresult}, we find out that with a significant level 5\%, the parameters $\xi_1$ of BTC is not significantly nonzero. In the remainder of this section, we will use the parameters in Table \ref{tab:mleresult} to carry out parameter estimations and the calculation of risk measures. 
\par
\begin{figure}[htbp!]
        \centering
        \subfigure[]{
        \includegraphics[width=4.5cm]{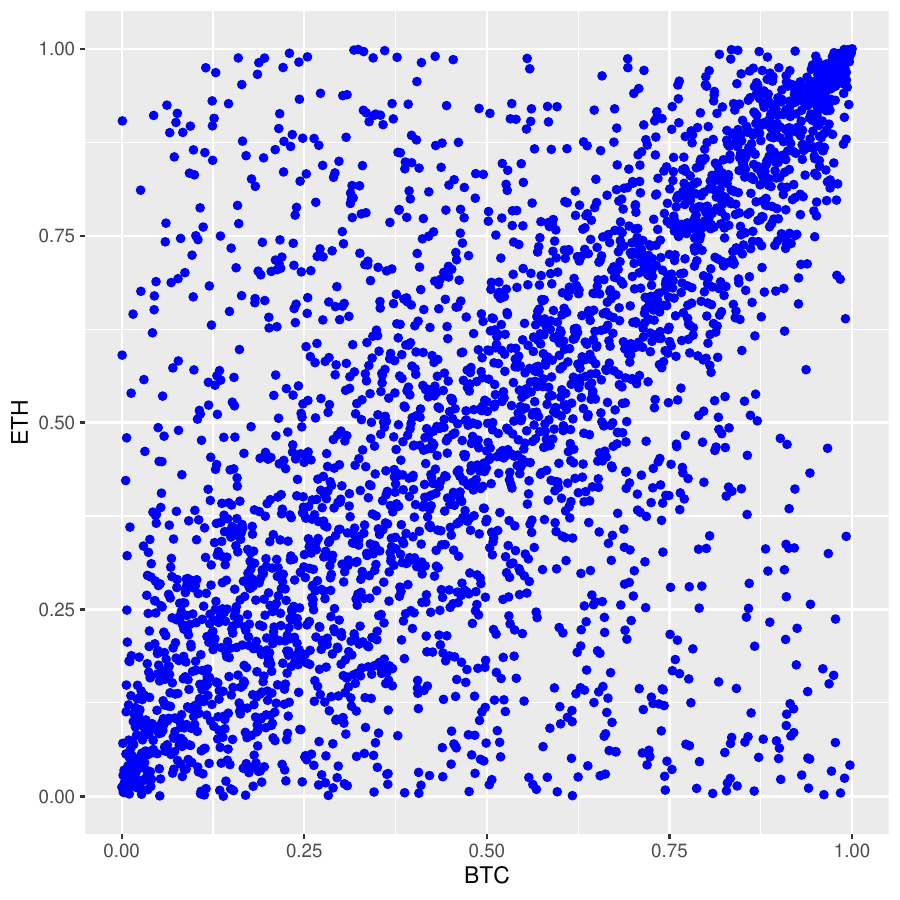}
        }
        \quad
        \subfigure[]{
        \includegraphics[width=4.5cm]{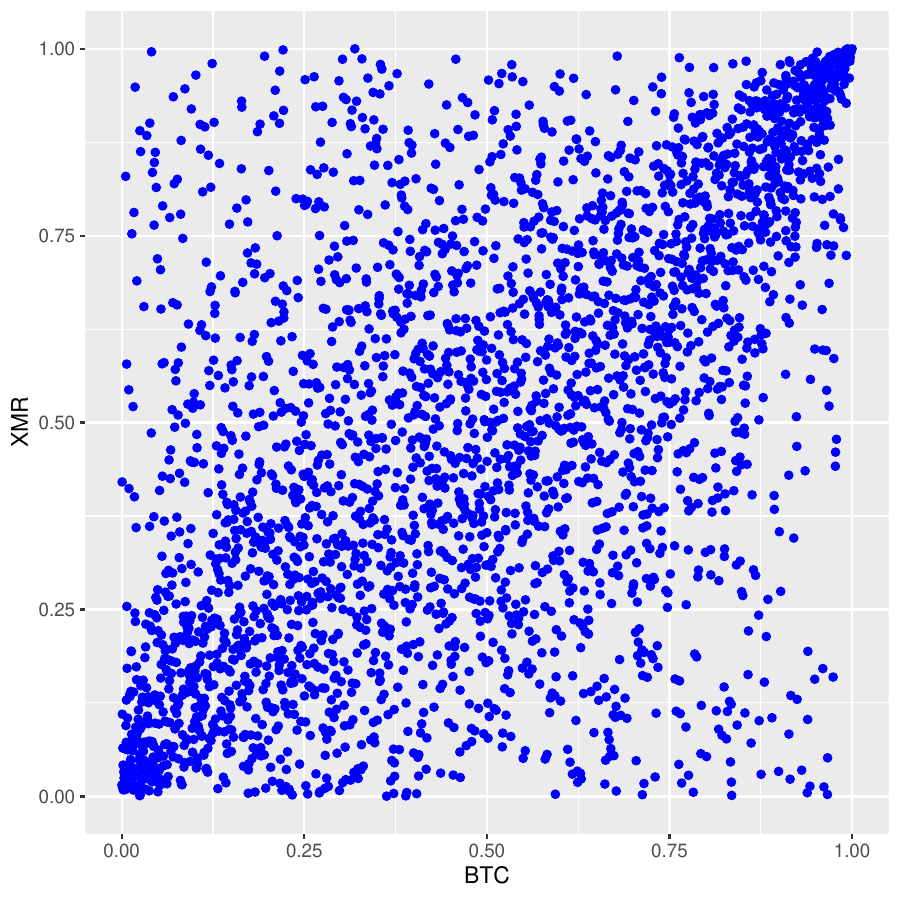}
        }
        \quad
        \subfigure[]{
        \includegraphics[width=4.5cm]{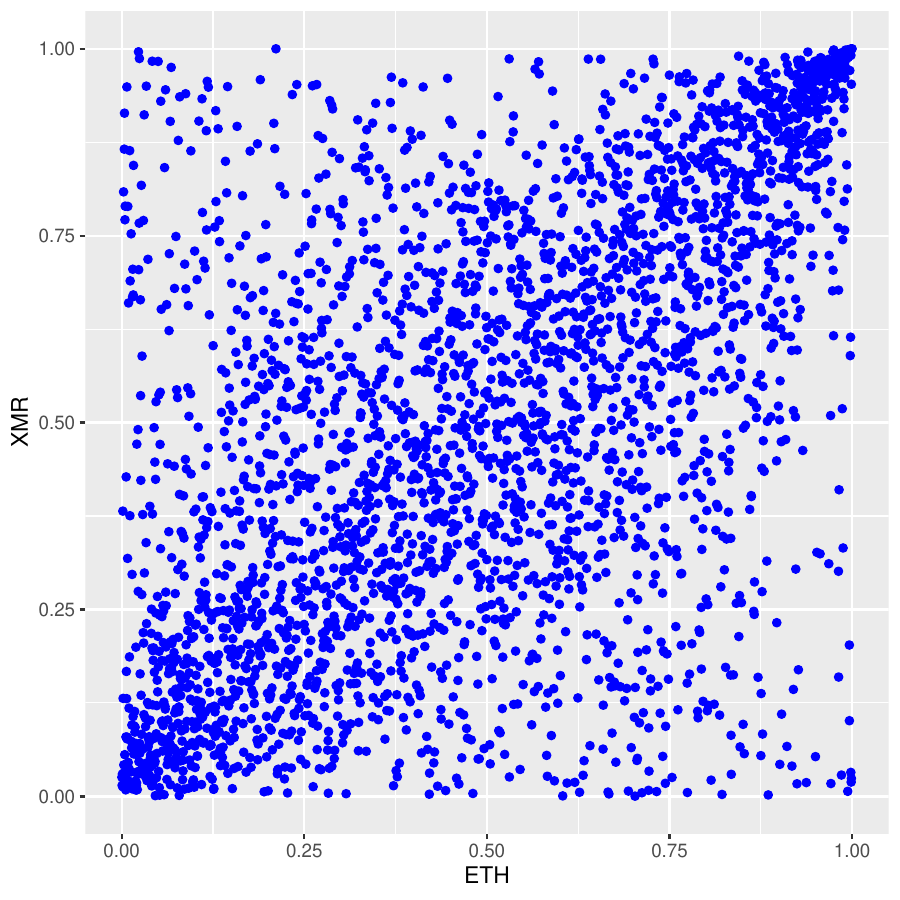}
        }
        \caption{Scatter plots of {$L_t$} for three paired cryptocurrencies.}
        \label{fig:Scatterplot}
        \end{figure}

 By employing the marginal distributions $\hat{F}_i$'s given in \eqref{eq:Fhatxin0} and \eqref{eq:Fhatxi0}, for each $i=1, \ldots, n$, we can convert observations $\left\{x_{i, k}\right\}_{1 \leq k \leq N}$ into pseudo-samples $\left\{\hat{U}_{i, k}\right\}_{1 \leq k \leq N}$, where
\begin{equation} \label{eq:psample}
    \hat{U}_{i, k}=\hat{F}_i\left(x_{i, k}\right), \quad 1 \leq k \leq N .
\end{equation}
Figure \ref{fig:Scatterplot} shows the scatter plot of pseudo-samples for different pairs of cryptocurrencies. 

\subsection{Parameter estimation of the mixed copula}
As per Table \ref{tab:corr}, it can be noted that the correlation coefficients between cryptocurrencies pairs are different, indicating that the interdependence among cryptocurrencies is significantly asymmetric. Hence, the dependence structure between these cryptocurrencies cannot be characterized by symmetric copulas. Besides, Figure \ref{fig:Scatterplot} shows that any pair of these three cryptocurrencies have a strong positive dependence at extreme values (for both of the left and right tails). Take these observations into consideration, we ought to use an asymmetric copula with non-zero left and right tail dependence coefficients to model the dependence structure between these cryptocurrencies. As a result, we employ a mixed copula model to characterize the dependence structure. The mixed copula model is widely used in the existing literature, like \cite{cai2014selection} and \cite{zhu2023asymptotic}.

\par We use the mixed Copula model
\begin{equation} \label{eq:mix-copula}
    \small
        C^M(\bm{u})=a_1 C^{Gau}(\bm{u}) + a_2 C^{Gum} (\bm{u}) + (1-a_1-a_2) C^{Cla}(\bm{u}), \quad 0 \leq a_1, a_2 \leq 1, \quad a_1 + a_2 \leq 1, \quad \bm{u} \in [0,1]^3
    \end{equation}
as the dependency model, \textcolor{blue}{where the Gaussian copula is used to capture the correlation between variables, the Gumbel copula models upper tail dependence, and the Clayton copula models lower tail dependence. We estimate these parameters using command ``\texttt{fitCopula}'' in R package ``\texttt{copula}''.} There are seven parameters $\left(\rho_{12}^M, \rho_{13}^M, \rho_{23}^M, \theta_G^M, \theta_C^M, a_1, a_2\right)$ to be estimated, where $\rho_{12}^M, \rho_{13}^M, \rho_{23}^M$ are the correlation coefficients in the Gaussian copula $C^{Gau}$, $\theta_G^M$ is the parameter in the Gumbel copula $C^{Gum}$, $\theta_C^M$ is the parameter in the Clayton copula $C^{Cla}$, and $a_1, a_2$ and $1-a_1-a_2$are the weights of the Gaussian copula, Gumbel copula, and Clayton copula, respectively. We performed maximum likelihood estimation on the pseudo-samples derived in \eqref{eq:psample} using seven different mixed Copula models and selected the most suitable model based on the Akaike Information Criterion (AIC) and Bayesian Information Criterion (BIC). The seven mixed Copula models are as follows: 
\begin{itemize}
    \item Model 1: Gaussian copula $C^{Gau}$;
    \item Model 2: Gumbel copula $C^{Gum}$;
    \item Model 3: Clayton copula $C^{Cla}$ ;
    \item Model 4: Gaussian-Gumbel mixed copula, formulated by setting $1-a_1 - a_2 = 0$ in \eqref{eq:mix-copula};
    \item Model 5: Gaussian-Clayton mixed copula, formulated by setting $a_2 = 0$ in \eqref{eq:mix-copula};
    \item Model 6: Gumbel-Clayton mixed copula, formulated by setting $a_1 = 0$ in \eqref{eq:mix-copula};
    \item Model 7: \textcolor{blue}{Gaussian-Gumbel-Clayton mixed copula}, formulated as in \eqref{eq:mix-copula}.
\end{itemize}
The mixed copula model with the smallest AIC and BIC values is selected as the most suitable model. The estimated parameters of the seven mixed copula models are shown in Table \ref{tab:ABIC}. The results in Tabel \ref{tab:ABIC} show that the mixed copula model with the smallest AIC and BIC values is Model 7, which is the \textcolor{blue}{Gaussian-Gumbel-Clayton mixed copula}. Then, we use the estimated parameters of Model 7 to calculate the risk measures in the following analysis. 
\begin{table}[]
 \renewcommand{\arraystretch}{1.35}
    \centering
    \caption{Performance of various mixed copula models}
    \label{tab:ABIC}
    \begin{tabular}{@{}clcc@{}}
    \toprule
                             & \multicolumn{1}{c}{Estimated Parameters}                                                        & AIC                        & BIC                        \\ \midrule
    Model 1                  & $\hat{\rho}_{12}^M = 0.630$, $\hat{\rho}_{13}^M = 0.560$, $\hat{\rho}_{23}^M = 0.567$ & -3349.372                  & -3331.136                  \\ \cline{2-4}
    Model 2                  & $\hat{\theta}_G^M = 1.727$                                                                      & -3911.606                  & -3905.527                  \\  \cline{2-4}
    Model 3                  & $\hat{\theta}_C^M = 0.806$                                                                      & -2202.380                  & -2196.302                  \\  \cline{2-4}
    \multirow{2}{*}{Model 4} & $\hat{\rho}_{12}^M = 0.140$, $\hat{\rho}_{13}^M = 0.095$, $\hat{\rho}_{23}^M = 0.076$,& \multirow{2}{*}{-4603.278} & \multirow{2}{*}{-4566.805} \\& $\hat{\theta}_G^M = 2.650$, $a_1 = 0.314$, $a_2 =0.686$ &                            &                            \\   \cline{2-4}
    \multirow{2}{*}{Model 5} & $\hat{\rho}_{12}^M = 0.5675$, $\hat{\rho}_{13}^M = 0.5288$, $\hat{\rho}_{23}^M = 0.4962$ & \multirow{2}{*}{-3608.101} & \multirow{2}{*}{-3571.629} \\ & $\hat{\theta}_G^M = 7.7034$, $a_1 = 0.8287$, $a_2 = 0$                                                                         &                            &                            \\    \cline{2-4}
    \multirow{2}{*}{Model 6} & $\hat{\theta}_G^M = 2.559$, $\hat{\theta}_C^M = 0.005 $  & \multirow{2}{*}{-4595.905} & \multirow{2}{*}{-4571.591} \\
                             & $a_1 = 0$, $a_2 =0.2770$                                                                         &                            &                            \\    \cline{2-4}
    \multirow{2}{*}{Model 7} & $\hat{\rho}_{12}^M = 0.104$, $\hat{\rho}_{13}^M = 0.061$, $\hat{\rho}_{23}^M = 0.048$, & \multirow{2}{*}{-4619.705} & \multirow{2}{*}{-4571.076} \\ & $\hat{\theta}_G^M = 2.583$, $\hat{\theta}_C^M = 9.465$, $a_1 = 0.299$, $a_2 =0.670 $                                                                         &                            &                            \\ \bottomrule
    \end{tabular}
    \end{table}

\subsection{Empirical results}
Based on the respective marginal distributions and the selected copula, a range of risk measures are provided in Table \ref{tab:measures}, where we take the confidence levels $p_1=p_2=p_3=0.95$\footnote{The market capitalization ratio among BTC, ETH, and XMR is approximately 9:3:1, considering the fluctuations in total market capitalization over time. We use this ratio as the reference weight for the calculations of MMME. Taking the MMME value of BTC as an example, the weight vector is set as $a= (0, 0.75, 0.25)$.}. More collected values under the settings of $p_1=p_2=p_3=0.975$ and $p_1=p_2=p_3=0.99$ can be found in Appendix \ref{appendix:tables}. The following observations can be noted\footnote{The specific definition s of $\Delta \mathrm{MCoVaR}$ and $\Delta ^{\mathrm{med}}\mathrm{MCoVaR}$ can be found in the appendix \ref{df:measures}.}:
\begin{table}[ht]
\centering
\caption{Values of some systemic risk measures of the three cryptocurrencies ($p_1=p_2=p_3=0.95$). }
\label{tab:measures}
\begin{tabular}{@{}cccc}
\toprule VaR-based & BTC & ETH & XMR \\
\midrule VaR & 5.727 & 7.892 & 8.084 \\
MCoVaR & 17.151 & 26.614 & 24.552 \\
$\Delta \mathrm{MCoVaR}$ & 11.424 & 18.722 & 16.468 \\
$\Delta ^{\mathrm{R}}\mathrm{MCoVaR}$ & 1.995 & 2.372 & 2.037 \\
$\Delta ^{\mathrm{med}}\mathrm{MCoVaR}$ & 8.951 & 15.140 & 13.202 \\
$\Delta ^{\mathrm{R-med}}\mathrm{MCoVaR}$ & 1.092 & 1.320 & 1.163 \\
\midrule ES-based & BTC & ETH & XMR \\
\midrule ES & 8.973 & 12.915 & 12.617 \\
MCoES & 21.427 & 35.450 & 31.652 \\
$\Delta \mathrm{MCoES}$ & 12.454 & 22.535 & 19.035 \\
$\Delta ^{\mathrm{R}}\mathrm{MCoES}$ & 1.388 & 1.745 & 1.509 \\
$\Delta ^{\mathrm{med}}\mathrm{MCoES}$ & 9.740 & 18.194 & 15.242 \\
$\Delta ^{\mathrm{R-med}}\mathrm{MCoES}$ & 0.833 & 1.054 & 0.929 \\
\midrule MMME-based& BTC & ETH & XMR \\
\midrule MMME & 2.166& 7.539&  6.798 \\
$\Delta \mathrm{MMME}$ & 2.082& 7.166& 6.457\\
$\Delta ^{\mathrm{R}}\mathrm{MMME}$ & 24.678& 19.240& 18.940 \\
\bottomrule &  &  &
\end{tabular}
\end{table}
\begin{enumerate}[(i)]
    \item Conditional risk measures, such as MCoVaR, MCoES and MMME, consistently surpass their unconditional counterparts, namely VaR and ES. This discrepancy underscores the heightened potential risk of individual assets when systemic risk is taken into account. These systemic risk measures capture the essence of how the performance of interconnected assets during periods of market extremity can amplify the risk profile of individual assets, thereby highlighting the risk co-movement effect. The implications are profound: there exists a pronounced risk interaction in the currency market, characterized by robust correlations among various currencies. Relying on unconditional risk measures in isolation may lead to an underestimation of the true risk exposure. 
    \item Under majority risk measures, ETH emerges with the highest risk profile, while BTC exhibits a comparatively lower risk level, with XMR occupying an intermediate position. This indicates that within the cryptocurrency market, ETH is more susceptible to systemic risk impacts, marked by notably higher price volatility and risk exposure in comparison to the other two digital assets. Conversely, BTC displays relative stability, with a lower risk exposure and spill-over effects, underscoring its status as the cryptocurrency with the largest market share and its perceived stability.
    \item Across a spectrum of risk measures, the risk ranking among the three currencies exhibits a noticeable consistency. This uniformity in risk perception across different cryptocurrencies indicates a stable market assessment of risk. It implies that within the currency market, relative measures such as ratio-based risk contribution measures (\(\Delta^{\rm R}\) and \(\Delta^{\rm R-med}\)) can effectively capture the systemic risk's relative change. These measures offer a nuanced approach beyond the reliance on absolute risk metrics like MCoVaR, MCoES and MMME.
    \item For the difference-based risk contribution measures (\(\Delta\) and \(\Delta^{\mathrm{med}}\)), the indicators derived from ES surpass those predicated on VaR. Conversely, for the ratio-based risk contribution measures (\(\Delta^{\mathrm{R}}\) and \(\Delta^{\mathrm{R-med}}\)), the scenario inverts, with VaR-based measures taking precedence. This suggests that ES-based measures confer a higher significance to individual assets' role in the distribution of systemic risk, while VaR-based measures accentuate the individual assets' proportional contribution to systemic risk as a whole. The disparity underscores the imperative to strike a balance in the selection of risk metrics, contingent upon the goals of risk management and the prevailing market conditions.
    \item The ratio-based risk contribution measures (\(\Delta^{\rm R}\) and \(\Delta^{\rm R-med}\)) offer a advantage over other risk metrics by virtue of their capacity to articulate the systemic risk's relative co-movement effect on individual assets via relative ratios. This methodology provides clarity on the comparative risk contributions of various assets within systemic risk frameworks. It affords a perspective that not only uncovers potential market extreme losses but also equips investors with a deeper comprehension and management of relative risk exposures across diverse market conditions. This is in contrast to the sole assessment of an asset's absolute risk level.

\end{enumerate}

\section{Conclusion} \label{sec:con}
Systemic risk plays a significant role in financial markets and portfolio management. This article delves into new tools for quantifying and analyzing systemic risk, with a specific emphasis on the absolute and relative spillover effects induced by systemic risk. Some comparison results are conducted based on these proposed measures for two different sets of multivariate vectors with the same or different copulas.
The theoretical findings have been validated through numerical examples, demonstrating the applicability and effectiveness of the proposed measures. Furthermore, we implement these measures as well as some known ones to quantify the interaction effect in  cryptocurrency market by considering three typical  cryptocurrencies.

\section*{Acknowledgements}
\textcolor{blue}{The authors are very grateful for the helpful comments and suggestions from two anonymous reviewers, which have improved the presentation of this manuscript.} Limin Wen thanks the financial support from the National Natural Science Foundation of China (No.72263019). Junxue Li thanks the Graduate Innovation Fund Project of Jiangxi Provincial Department of Education (No. YC2024-B089). Yiying Zhang acknowledges the financial support from the GuangDong Basic and Applied Basic Research Foundation (No. 2023A1515011806), and Shenzhen Science and Technology Program (No. JCYJ20230807093312026).


\section*{Disclosure statements}
No potential competing interest was reported by the authors.


\appendix

\bibliographystyle{mystyle}
\setlength{\bibsep}{0.0pt}
{\small\bibliography{reference}}

\appendix

\section{Proof of the main results} \label{app:A}
For a random variable $X$ with distribution function $F_X$, if $h$ is continuous, then $h\left(\overline F_X(x)\right) = h \circ \overline F_X(x)$ applied to the tail function $\overline F_X (x) = 1- F_X(x)$ results in a new tail function. This new tail function corresponds to a random variable  $X_h$, which is derived from $X$ by applying the distortion function $h$. The following lemma derives the distorted function of $X_1$ given that the remaining entities are in distress, as discussed in \citet{sordo2015comparison} and \citet{ortega2021}.

\begin{Lem}\label{Lem_distortion}
Let $\bm X = (X_1,\dots,X_n)$ be an $n$-dimensional random vector with copula $C$, joint distribution function $F$ and marginal distributions $F_1,\dots,F_n$. Assume that $(X_2,\dots,X_n) \uparrow_{\rm SI} X_1$. Then, for $(p_2,\dots,p_n) \in (0,1)^{n-1}$, the conditional random variable $\left[ X_1 \bigg| \bigcap\limits_{j = 2}^n \left\{{X_j} > {\rm VaR}_{p_j}[X_j]\right\}  \right]$ is a distorted random variable induced from $X_1$ by the concave distortion function
\begin{equation}
    h_{\bm p}(t) = \frac{\overline C (1-t,p_2,\dots,p_n)}{\overline C (0,p_2,\dots,p_n)},~~t \in [0,1].
\end{equation}   
\end{Lem}

 \subsection{Proof of Theorem \ref{The_RCoVaR}}
\begin{proof}
 Let \( X_{h,\bm p} \) be defined as the conditional random variable \( \left[ X_1 \bigg| \bigcap\limits_{j = 2}^n \left\{{X_j} > {\rm VaR}_{p_j}[X_j]\right\} \right] \) for \( \bm p = (p_1,...,p_n) \in (0,1)^n \). Assuming \( X_{h,\bm p} \sim F_{X_{h,\bm p}} \), it follows that 
\[
{\rm MCoVaR}_{\bm p}[X_1|X_2,...,X_n] = F_{X_{h,\bm p}}^{-1}(p_1).
\]
By applying Lemma \ref{Lem_distortion}, we find that $F_{X_{h,{\bm p}}}(x) = 1 - h_{\bm p}\left(\overline{F}_1(x)\right)$,
where \( h_{\bm p}(t) \) is given by 
\[
h_{\bm p}(t) = \frac{\overline{C}(1-t,p_2,...,p_n)}{\overline{C}(0,p_2,...,p_n)}.
\]
Setting \( F_{X_{h,{\bm p}}}(x) = 1 - h_{\bm p}(\overline{F}_1(x)) = p_1 \), we obtain $F_{X_{h, {\bm p}}}^{-1}(p_1) = F_1^{-1}\left(1 - h_{\bm p}^{-1}(1-p_1)\right)$.  Hence, we have
\begin{equation*}
     \Delta^{\rm R} {\rm MCoVaR}_{\bm p}[X_1|X_2,\dots,X_n] = \frac{F_1^{-1}\left(1-h_{\bm p}^{-1}(1-p_1)\right)}{F_1^{-1}(p_1)}-1.
\end{equation*}
Similarly,
    \begin{equation*}
     \Delta^{\rm R} {\rm MCoVaR}_{\bm p}[Y_1|Y_2,\dots,Y_n] = \frac{G_1^{-1}\left(1-{h'}_{\bm p}^{-1}(1-p_1)\right)}{G_1^{-1}(p_1)}-1,
\end{equation*}
where $h'_{\bm p}(t) = \frac{\overline C'(1-t,p_2,\dots,p_n)}{\overline C'(0,p_2,\dots,p_n)}$.  
Without loss of generality, we assume that $(X_2,..,X_n) \uparrow_{\rm SI} X_1$. The proof for the other case is similar.  Lemma \ref{Lem_distortion} indicates that \( h_{\bm p} \) is a concave distortion function, satisfying \( h_{\bm p}(t) \geq t \), which leads to \( t \leq 1-h_{\bm p}^{-1}(1-t) \).
By applying \( X_1 \leq_{\star} Y_1 \), it follows that
$$\frac{F_1^{-1}\left(1-h_{\bm p}^{-1}(1-p_1)\right)}{F_1^{-1}(p_1)} \leq \frac{G_1^{-1}\left(1-h_{\bm p}^{-1}(1-p_1)\right)}{G_1^{-1}(p_1)}.$$
Utilizing the definition of \( C \leq_{\rm whr} C' \), we directly obtain that  
\[  \frac{\overline {C}'(0,p_2,\dots,p_n)}{\overline C(0,p_2,\dots,p_n)} \leq \frac{\overline {C}'(1-t,p_2,\dots,p_n)}{\overline C(1-t,p_2,\dots,p_n)},  \]
which leads to \( h_{\bm p}(t) \leq h'_{\bm p}(t) \) for all \( t \in (0,1) \).
Hence, $G_1^{-1}\left(1-h_{\bm p}^{-1}(1-p_1)\right) \leq G_1^{-1} \left(1-{h'}_{\bm p}^{-1}(1-p_1)\right)$, confirming that (\ref{eq_RMCoVaR}) is valid. This completes the proof.
\end{proof}

 \subsection{Proof of Theorem \ref{The_RCoES}}
\begin{proof}
 Given Definition \ref{def_RMCoES} and the proof of Theorem \ref{The_RCoVaR}, the MCoES can be reformulated as:
\begin{equation*}
    {\rm MCoES}_{\bm p}[X_1|X_2,...,X_n] = \frac{1}{1-p_1} {\int_{p_1}^1 F_1^{-1}\left(1 - h_{\bm p}^{-1} (1-t)\right) dt} = \int_0^1 F_1^{-1}(s) d A_{\bm p}(s), 
\end{equation*}
where \( A_{\bm p}(s) \) is specified as
\[
A_{\bm p}(s) = 
\begin{cases} 
0, & s \leq 1 - h_{\bm p}^{-1} (1-p_1), \\
1-\frac{1}{1-p_1} \cdot \frac{\overline{C}(s,p_2,...,p_n)}{\overline{C}(0,p_2,...,p_n)}, & s > 1 - h_{\bm p}^{-1} (1-p_1),
\end{cases}
\]
for \( \bm p \in (0,1)^n \). Hence, it follows that
\begin{equation*}
     \Delta^{\rm R} {\rm MCoES}_{\bm p}[X_1|X_2,\dots,X_n] = \frac{\int_0^1 F_1^{-1}(s) dA_{\bm p}(s)}{\int_0^1 F_1^{-1}(s) dB(s)}-1,
\end{equation*}
where
\[
A_{\bm p}(s) =
\begin{cases} 
0, & s \leq 1 - h_{\bm p}^{-1} (1-p_1), \\
1-\frac{1}{1-p_1} \cdot \frac{\overline{C}(s,p_2,\dots,p_n)}{\overline{C}(0,p_2,\dots,p_n)}, &  s > 1 - h_{\bm p}^{-1} (1-p_1),
\end{cases}
\]
for $\bm p\in (0,1)^n$, with $h_{\bm p}(t) = \frac{\overline C(1-t,p_2,\dots,p_n)}{\overline C(0,p_2,\dots,p_n)}$ and $B(s) = {\rm max} \left\{0, \frac{s-p_1}{1-p_1} \right\}$.
Similarly,
\begin{equation*}
     \Delta^{\rm R} {\rm MCoES}_{\bm p}[Y_1|Y_2,\dots,Y_n] = \frac{\int_0^1 G_1^{-1}(s) dA'_{\bm p}(s)}{\int_0^1 G_1^{-1}(s) dB(s)}-1,
\end{equation*}
where
\[
A'_{\bm p}(s) =
\begin{cases} 
0, & s \leq 1 - {h}_{\bm p}'^{-1} (1-p_1), \\
1-\frac{1}{1-p_1} \cdot \frac{\overline{C}'(s,p_2,\dots,p_n)}{\overline{C}'(0,p_2,\dots,p_n)}, &  s > 1 - {h}_{\bm p}'^{-1} (1-p_1),
\end{cases}
\]
for $\bm p\in (0,1)^n$, with \( h'_{\bm p}(t) = \frac{\overline C'(1-t,p_2,\dots,p_n)}{\overline C'(0,p_2,\dots,p_n)} \). Since \( (X_2,\dots,X_n) \uparrow_{\rm SI} X_1 \), it can be inferred from Lemma \ref{Lem_distortion} that \( A_{\bm p}(s) \) is convex, and thus \( A_{\bm p}\left(B^{-1}(s)\right) \) is also convex in \( s \). By Lemma \ref{Lem_belzunce}, it follows that 
\begin{equation*}
    \frac{\int_0^1 F_1^{-1}(s) dA_{\bm p}(s)}{\int_0^1 F_1^{-1}(s) dB(s)}-1 \leq \frac{\int_0^1 G_1^{-1}(s) dA_{\bm p}(s)}{\int_0^1 G_1^{-1}(s) dB(s)}-1.
\end{equation*}
Besides, the condition \( C \leq_{\rm whr} C' \) ensures that \( h_{\bm p}(t) \leq h'_{\bm p}(t) \), which further implies \( A_{\bm p}(s) \geq A'_{\bm p}(s) \). Using integration by parts, we obtain
\begin{equation*}
    \int_0^1 G_1^{-1}(s) dA_{\bm p}(s) - \int_0^1 G_1^{-1}(s) dA'_{\bm p}(s) = \int_0^1 \left(A_{\bm p}(s) - A'_{\bm p}(s) \right) d G_1^{-1}(s) \leq 0.
\end{equation*}
which establishes (\ref{eq_RMCoES}) and completes the proof.
\end{proof}

\subsection{Proof of Theorem \ref{The_RMMMES}}
\begin{proof}
Define $U_i = F_i(X_i)$ for $i=1,...,n$. The MMME is then expressed as 
\begin{eqnarray*}
    &&  {\rm MMME}_{\bm p_{[-1]}}[X_{1}|X_{2},...,X_{n}] = \mathbb E\left[ (X_1-A_{X,\bm p_{[-1]}})_+ \bigg| \bigcap\limits_{j = 2}^n \left\{{X_j} > {\rm VaR}_{p_j}[X_j]\right\} \right] \\
    &=& \int_{A_{X,\bm p_{[-1]}}}^\infty \left( 1- F_{U_1 | \bigcap\limits_{j = 2}^n U_j > p_j}(F_1(t)) \right) dt = \int_{F_1\left(A_{X,\bm p_{[-1]}}\right)}^1 \left( 1- F_{U_1 | \bigcap\limits_{j = 2}^n U_j > p_j}(u) \right) d F_1^{-1}(u) \\
    &=& \int_{F_1\left(A_{X,\bm p_{[-1]}}\right)}^1 \left( F_1^{-1}(u) - A_{X,\bm p_{[-1]}} \right) d \overline h_{\bm p}(u),
\end{eqnarray*}
where $\overline h_{\bm p}(t) = 1- h_{\bm p}(1-t)$. We have
\begin{equation*}
     \Delta^{\rm R} \mathrm{MMME}_{\bm p_{[-1]}}[X_{1}|X_{2},\dots,X_{n}] = \frac{\int_{F_1\left(A_{X,\bm p_{[-1]}}\right)}^1 \left( F_1^{-1}(u) - A_{X,\bm p_{[-1]}} \right) d \overline h_{\bm p}(u)}{\int_{F_1\left(A_{X,\bm p_{[-1]}}\right)}^1 \left( F_1^{-1}(u) - A_{X,\bm p_{[-1]}} \right) du} -1,
\end{equation*}
where $h_{\bm p}(t) = \frac{\overline C(1-t,p_2,\dots,p_n)}{\overline C(0,p_2,\dots,p_n)}$ and $\overline h_{\bm p}(t) = 1- h_{\bm p}(1-t)$. Similarly,
\begin{equation*}
     \Delta^{\rm R} \mathrm{MMME}_{\bm p_{[-1]}}[Y_{1}|Y_{2},\dots,Y_{n}] = \frac{\int_{G_1\left(A_{Y,\bm p_{[-1]}}\right)}^1 \left( G_1^{-1}(u) - A_{Y,\bm p_{[-1]}} \right) d \overline h'_{\bm p}(u)}{\int_{G_1\left(A_{Y,\bm p_{[-1]}}\right)}^1 \left( G_1^{-1}(u) - A_{Y,\bm p_{[-1]}} \right) du} -1,
\end{equation*}
where $h'_{\bm p}(t) = \frac{\overline C'(1-t,p_2,\dots,p_n)}{\overline C'(0,p_2,\dots,p_n)}$ and $\overline {h}'_{\bm p}(t) = 1- h'_{\bm p}(1-t)$.  
Given that \( X_i =_{\rm st} Y_i \), it suffices to demonstrate that
$$\int_{F_1\left(A_{X,\bm p_{[-1]}}\right)}^1 \left( F_1^{-1}(u) - A_{X,\bm p_{[-1]}} \right) d \overline h_{\bm p}(u) \leq \int_{F_1\left(A_{X,\bm p_{[-1]}}\right)}^1 \left( F_1^{-1}(u) - A_{X,\bm p_{[-1]}} \right) d \overline h'_{\bm p}(u),$$
which follows from \( C \leq_{\rm whr} C' \). This completes the proof.
\end{proof}

 \subsection{Proof of Theorem \ref{The_RmedCoVaR}}
\begin{proof}
By the proof of Theorem \ref{The_RCoVaR}, it follows that
\begin{equation*}
        \Delta^{\rm R-med} {\rm MCoVaR}_{\bm p}(X_1|X_2,\dots,X_n)  = \frac{F_1^{-1}\left(1-h_{\bm p_{[-1]}}^{-1}(1-p_1)\right)}{F_1^{-1}\left(1-h_{\frac{\bm 1}{\bm 2}}^{-1}(1-p_1)\right)}-1,
    \end{equation*}
    where $h_{\hat {\bm p}}(t) = \frac{\overline C(1-t,p_2,\dots,p_n)}{\overline C(0,p_2,\dots,p_n)}$ with $\hat {\bm p} =\left\{\bm p_{[-1]}, \frac{\bm 1}{\bm 2}\right\}$, $\bm p_{[-1]} = (p_2,\dots,p_n) \in (1/2,1)^{n-1}$ and $\frac{\bm 1}{\bm 2} = \left(\frac{1}{2},\dots,\frac{1}{2}\right) \in \mathbb R^{n-1}$. Similarly,
    \begin{equation*}
        \Delta^{\rm R-med} {\rm MCoVaR}_{\bm p}[Y_1|Y_2,\dots,Y_n]  = \frac{G_1^{-1}\left(1-h_{\bm p_{[-1]}}^{-1}(1-p_1)\right)}{G_1^{-1}\left(1-h_{\frac{\bm 1}{\bm 2}}^{-1}(1-p_1)\right)}-1.
    \end{equation*}
Given that \( X_1 \uparrow_{\rm RTI} (X_2,\dots,X_n) \), it follows that \( h_{\bm p_{[-1]}}(t) \geq h_{\frac{\bm 1}{\bm 2}}(t) \), which implies \( 1-h_{\frac{\bm 1}{\bm 2}}^{-1}(1-p_1) \leq 1-h_{\bm p_{[-1]}}^{-1}(1-p_1) \). Hence, \( X_1 \leq_{\star} Y_1 \) implies that
\begin{equation*}
    \frac{G_1^{-1}\left(1-h_{\frac{\bm 1}{\bm 2}}^{-1}(1-p_1)\right)}{F_1^{-1}\left(1-h_{\frac{\bm 1}{\bm 2}}^{-1}(1-p_1)\right)} \leq \frac{G_1^{-1}\left(1-h_{\bm p_{[-1]}}^{-1}(1-p_1)\right)}{F_1^{-1}\left(1-h_{\bm p_{[-1]}}^{-1}(1-p_1)\right)},
\end{equation*}
which confirms that (\ref{eq_RmedMCoVaR}) is satisfied. This completes the proof.
\end{proof}

 \subsection{Proof of Theorem \ref{The_RmedCoES}}
\begin{proof}
By the proof of Theorem \ref{The_RCoES},  we obtain
\begin{equation*}
   \Delta^{\rm R-med} {\rm MCoES}_{\bm p}[X_1|X_2,\dots,X_n] =  \frac{\int_0^1 F_1^{-1}(s) dA_{\bm p_{[-1]}}(s)}{\int_0^1 F_1^{-1}(s) d A_{\frac{\bm 1}{\bm 2}}(s)}-1
\end{equation*}
and
\begin{equation*}
   \Delta^{\rm R-med} {\rm MCoES}_{\bm p}[Y_1|Y_2,\dots,Y_n] =  \frac{\int_0^1 G_1^{-1}(s) dA_{\bm p_{[-1]}}(s)}{\int_0^1 G_1^{-1}(s) d A_{\frac{\bm 1}{\bm 2}}(s)}-1,
\end{equation*}
where
\[
A_{\hat {\bm p}}(s) =
\begin{cases} 
0, & s \leq 1 - h_{\hat {\bm p}}^{-1} (1-p_1), \\
1-\frac{1}{1-p_1} \cdot \frac{\overline{C}(s,p_2,\dots,p_n)}{\overline{C}(0,p_2,\dots,p_n)}, & s > 1 - h_{\hat {\bm p}}^{-1} (1-p_1),
\end{cases}
\]
for \( p_1 \in (0,1) \) and \( \hat {\bm p} \in \left\{\bm p_{[-1]}, \frac{\bm 1}{\bm 2} \right\} \), with \( \bm p_{[-1]} = (p_2,\dots,p_n) \in (1/2,1)^{n-1} \) and \( \frac{\bm 1}{\bm 2} = \left(\frac{1}{2},\dots,\frac{1}{2}\right) \in \mathbb R^{n-1} \).   \\ 
Given that \( C \) is \( {\rm MTP}_2 \), and \( (X_2,\dots,X_n) \uparrow_{\rm SI} X_1 \), Lemma \ref{Lem_distortion} implies that \( h_{\hat {\bm p}} \) is a concave distortion function, making \( A_{\hat {\bm p}}(s) \) a convex distortion function.
To apply Lemma \ref{Lem_belzunce}, we demonstrate that \( A_{\bm p_{[-1]}} \left(A_{\frac{\bm 1}{\bm 2}}^{-1}(s) \right) \) is convex, equivalent to showing that \( \left(A_{\bm p_{[-1]}} (s) \right)' / \left(A_{\frac{\bm 1}{\bm 2}}(s) \right)' \) is increasing in \( s \), where
\begin{equation*}
    \left(A_{\hat {\bm p}} (s)\right)' = \frac{\partial A_{\hat {\bm p}}(s)}{\partial s} = \frac{1}{1-p_1} \cdot \frac{\mathbb P(U_2>p_2,\dots,U_n>p_n|U_1=s)}{\overline{C}(0,p_2,\dots,p_n)},
\end{equation*}
with \( U_i = F_i(X_i) \). Using \( C \) being \( {\rm MTP}_2 \), we have
\[
(U_2,\dots,U_n|U_1 = u_1) \leq_{\rm whr} (U_2,\dots,U_n|U_1 = u_1'), \quad \forall u_1 \leq u_1',
\]
which implies that
\[
\frac{\left(A_{\bm p_{[-1]}} (s) \right)'}{\left(A_{\frac{\bm 1}{\bm 2}}(s) \right)'} = \frac{\overline{C}(0,1/2,\dots,1/2)}{\overline{C}(0,p_2,\dots,p_n)} \cdot \frac{\mathbb P(U_2>p_2,\dots,U_n>p_n|U_1=s)}{\mathbb P(U_2>1/2,\dots,U_n>1/2|U_1=s)}
\]
is increasing in \( s \). 
It has been shown that \( A_{\bm p_{[-1]}}(s) \), \( A_{\frac{\bm 1}{\bm 2}}(s) \), and \( A_{\bm p_{[-1]}}\left(A^{-1}_{\frac{\bm 1}{\bm 2}}(s)\right) \) are convex functions. Therefore, applying Lemma \ref{Lem_belzunce} and  \( X_1 \leq_{\rm ps} Y_1 \), we deduce that (\ref{eq_RmedMCoES}) holds. This completes the proof.
\end{proof}

\section{Supplementary materials}

\subsection{Risk measures} \label{df:measures}
The contribution measures employed in Tables \ref{tab:measures} are presented, as detailed in \cite{sordo2018} and \cite{ortega2021}.
\begin{Def} \label{def_MCoVaR}
For any $p_1 \in (0,1)$, the difference-based contribution $\rm MCoVaR$ with systemic risk event  is defined by
\begin{equation}
    \Delta {\rm MCoVaR}_{\bm p}[X_{1}|X_{2},\dots,X_{n}] = {\rm MCoVaR}_{p_1, \bm p_{[-1]}}[X_{1}|X_{2},\dots,X_{n}] -{\rm VaR}_{p_1}[X_1],
\end{equation}
and
\begin{equation}
    \Delta^{\rm med} {\rm MCoVaR}_{\bm p}[X_{1}|X_{2},\dots,X_{n}] = {\rm MCoVaR}_{p_1, \bm p_{[-1]}}[X_{1}|X_{2},\dots,X_{n}] - {\rm MCoVaR}_{ p_1,\frac{\bm 1}{\bm 2}}[X_{1}|X_{2},\dots,X_{n}],
\end{equation}
where $\bm p_{[-1]} = (p_2,\dots,p_n) \in (1/2,1)^{n-1}$ and $\frac{\bm 1}{\bm 2} = \left(\frac{1}{2},\dots,\frac{1}{2}\right) \in \mathbb R^{n-1}$.
\end{Def}

\begin{Def}
    For $p_1 \in (0,1)$, the difference-based contribution $\rm MCoES$ with systemic risk event  is defined by
\begin{equation}
    \Delta {\rm MCoES}_{\bm p}[X_{1}|X_{2},\dots,X_{n}] = {\rm MCoES}_{p_1, \bm p_{[-1]}}[X_{1}|X_{2},\dots,X_{n}] -{\rm ES}_{p_1}[X_1],
\end{equation}
and
\begin{equation}
    \Delta^{\rm med} {\rm MCoES}_{\bm p}[X_{1}|X_{2},\dots,X_{n}] = {\rm MCoES}_{p_1, \bm p_{[-1]}}[X_{1}|X_{2},\dots,X_{n}] - {\rm MCoES}_{ p_1,\frac{\bm 1}{\bm 2}}[X_{1}|X_{2},\dots,X_{n}],
\end{equation}
where $\bm p_{[-1]} = (p_2,\dots,p_n) \in (1/2,1)^{n-1}$ and $\frac{\bm 1}{\bm 2} = \left(\frac{1}{2},\dots,\frac{1}{2}\right) \in \mathbb R^{n-1}$.
\end{Def}

\subsection{Copula} \label{app:copula}
\begin{Def}
    Setting the generating function to $\psi (u) = (-{\rm ln}~u)^\theta$, thus $\psi^{-1}(u) = {\rm exp} \left(-u^{\frac{1}{\theta}}\right)$. The $n$-dimensional Gumbel copula is defined as follows: 
\begin{equation*}
    C_\theta(u_1,\dots,u_n) = {\rm exp}\left\{ -\left[ \sum_{i=1}^n (-{\rm ln}~ u_i)^\theta \right]^\frac{1}{\theta}  \right\},~
    \theta>1,~{\bm u} \in [0,1]^n.
\end{equation*}
\end{Def}
Gumbel copula exhibits different dependency properties in the left and right tails. Typically, it demonstrates positive right-tail dependency, implying that when one variable exhibits an extreme value in the right tail, there is a higher probability for the other variable to also have an extreme value in the right tail.

\begin{Def}
    Setting the generating function to $\psi (u) = u^{-\theta} - 1$, thus $\psi^{-1}(u) = (u+1)^{\frac{1}{\theta}}$. The $n$-dimensional Clayton copula is defined as follows: 
\begin{equation*}
    C_\theta(u_1,\dots,u_n) =  \left[ \sum_{i=1}^n u_i^{-\theta} -n +1 \right]^{-\frac{1}{\theta}},~
    \theta>0,~{\bm u} \in [0,1]^n.
\end{equation*}
\end{Def}
The Clayton copula exhibits significant dependency in the left tail, meaning that when one variable exhibits an extreme value in the left tail, there is a higher probability for the other variable to also have an extreme value in the left tail. Therefore, the combination of the Gumbel copula and the Clayton copula can simulate asymmetric upper and lower tail dependencies.

In addition to Archimedean copulas, there is another class of copula functions called elliptical copulas, such as the Gaussian copula. The Gaussian copula exhibits a certain degree of symmetry in terms of its dependence properties in the left and right tails. which is defined as follows. 
\begin{Def}
Let $R$ be a symmetric, positive definite matrix with ${\rm diag}(R) = (1,\dots,1)'$ and $\Phi_R$ the standardized multivariate normal distribution with correlation matrix  $R$. The multivariate Gaussian copula is defined as follows:
\begin{equation*}
    C_R(u_1,\dots,u_n) = \Phi_R\left( \Phi^{-1}(u_1),\dots,\Phi^{-1}(u_n) \right),
\end{equation*}
where $\Phi^{-1}$ is the inverse of the standard univariate normal distribution function $\Phi$.
\end{Def}

\subsection{Tail dependence coefficient}
The tail dependence coefficient is a measure of the dependence between random variables in the tails of their joint distribution. It is divided into the upper tail dependence coefficient and the lower tail dependence coefficient, which describe the dependence of random variables in the upper and lower tails of their joint distribution, respectively. Based on this concept, the notion of multivariate upper and lower tail dependence coefficients is introduced.
\begin{Def}
    For a random vector \(\bm{X} = (X_1, \ldots, X_n)\), let \(S\) be a randomly chosen subset of \(\{1, \ldots, n\}\) with \(|S| = k\), and let \(\bar{S} = \{1, \ldots, n\} \setminus S\). The multivariate upper tail dependence coefficient is defined as:
\begin{equation*}
    \lambda_U^{S|\bar{S}} = \lim_{u \to 1^{-}} \mathbb{P}\left(\bigcap_{i \in S} \{F_{i}(X_i) > u\} \bigg | \bigcap_{j \in \bar{S}} \{F_{j}(X_j) > u\}\right).
\end{equation*}
The multivariate lower tail dependence coefficient is defined as:
\begin{equation*}
    \lambda_L^{S|\bar{S}} = \lim_{u \to 0^{+}} \mathbb{P}\left(\bigcap_{i \in S} \{F_{i}(X_i) \leq u\} \bigg| \bigcap_{j \in \bar{S}} \{F_{j}(X_j) \leq u\}\right).
\end{equation*}
\end{Def}

\subsection{Additional tables under different confidence levels}\label{appendix:tables}
Tables \ref{tab:measures_0.975} and \ref{tab:measures_0.99} summarize the computed values of some systemic risk measures of the three cryptocurrencies under $p_1=p_2=p_3=0.975$ and $p_1=p_2=p_3=0.99$, respectively. It can be noted that the overall trend of these  risk measures in the two additional tables is similar to the one  for \( p_1 = p_2 = p_3 = 0.95 \). Moreover, as \( p_i \) (\( i = 1, 2, 3 \)) increases, the risk measures generally show an upward trend. This indicates that as the extremity of the conditional events increases, the value of co-risk measures also grows.

\begin{table}[!htbp]
    \centering
    \begin{minipage}{.45\textwidth}
      \centering
      \captionof{table}{Values of some systemic risk measures of the three cryptocurrencies ($p_1=p_2=p_3=0.975$). }
      \label{tab:measures_0.975}
      \begin{tabular}{@{}cccc}
\toprule VaR-based & BTC & ETH & XMR \\
\midrule VaR & 7.874
& 11.003
& 10.968 \\
MCoVaR & 23.078 & 38.473& 34.230\\
$\Delta \mathrm{MCoVaR}$ & 15.205& 27.470& 23.262 \\
$\Delta ^{\mathrm{R}}\mathrm{MCoVaR}$ & 1.931& 2.497& 2.121\\
$\Delta ^{\mathrm{med}}\mathrm{MCoVaR}$ & 12.571& 23.417& 19.659\\
$\Delta ^{\mathrm{R-med}}\mathrm{MCoVaR}$ & 1.196& 1.555& 1.349\\
\midrule ES-based & BTC & ETH & XMR \\
\midrule ES & 11.313& 16.657& 15.948\\
MCoES & 27.880& 49.701& 42.822\\
$\Delta \mathrm{MCoES}$ & 16.567& 33.044& 26.873\\
$\Delta ^{\mathrm{R}}\mathrm{MCoES}$ & 1.464& 1.984& 1.685\\
$\Delta ^{\mathrm{med}}\mathrm{MCoES}$ & 13.680& 28.139& 22.694\\
$\Delta ^{\mathrm{R-med}}\mathrm{MCoES}$ & 0.963& 1.305& 1.127\\
\midrule MMME-based& BTC & ETH & XMR \\
\midrule MMME & 1.943& 8.100&  7.195\\
$\Delta \mathrm{MMME}$ & 1.907 & 7.913& 7.025\\
$\Delta ^{\mathrm{R}}\mathrm{MMME}$ &  52.467& 42.147& 41.358 \\
\bottomrule &  &  &
\end{tabular}
    \end{minipage}%
    \hspace{0.05\textwidth}
    \begin{minipage}{.45\textwidth}
      \centering
      \captionof{table}{Values of some systemic risk measures of the three cryptocurrencies ($p_1=p_2=p_3=0.99$). }
      \label{tab:measures_0.99}
      \begin{tabular}{@{}cccc}
\toprule VaR-based & BTC & ETH & XMR \\
\midrule VaR & 10.908& 15.723& 15.222\\
MCoVaR & 31.983& 59.056& 50.088\\
$\Delta \mathrm{MCoVaR}$ & 21.075& 43.333& 34.865\\
$\Delta ^{\mathrm{R}}\mathrm{MCoVaR}$ & 1.932& 2.756& 2.290\\
$\Delta ^{\mathrm{med}}\mathrm{MCoVaR}$ & 18.217& 38.570& 30.770\\
$\Delta ^{\mathrm{R-med}}\mathrm{MCoVaR}$ & 1.323& 1.883& 1.593\\
\midrule ES-based & BTC & ETH & XMR \\
\midrule ES & 14.618& 22.333& 20.862\\
MCoES & 37.578& 74.451& 61.134\\
$\Delta \mathrm{MCoES}$ & 22.960& 52.117& 40.272\\
$\Delta ^{\mathrm{R}}\mathrm{MCoES}$ & 1.571& 2.334& 1.930\\
$\Delta ^{\mathrm{med}}\mathrm{MCoES}$ & 19.829& 46.355& 35.523\\
$\Delta ^{\mathrm{R-med}}\mathrm{MCoES}$ & 1.117& 1.650& 1.387\\
\midrule MMME-based & BTC & ETH & XMR \\
\midrule MMME & 1.619  & 10.394& 8.464\\
$\Delta \mathrm{MMME}$ & 1.607& 10.294& 8.385\\
$\Delta ^{\mathrm{R}}\mathrm{MMME}$ & 137.496& 103.021 & 105.524 \\
\bottomrule &  &  &
\end{tabular}
\end{minipage}
\end{table}

\end{document}